\documentclass[10pt,a4paper]{jpconf}
\usepackage{subcaption} 

\usepackage{fancyhdr}

\usepackage{titling}

\usepackage[normalem]{ulem}
\usepackage{color,soul}
\usepackage{amsmath}
\usepackage{graphicx}
\usepackage[a4paper]{geometry}
\usepackage{babel}
\usepackage{makeidx} 
\usepackage{multicol}
\usepackage[T1]{fontenc}
\usepackage{float}
\usepackage{rotating}
\usepackage[export]{adjustbox}
\usepackage{textcomp}
\usepackage{color,soul}
\usepackage{marvosym}

\usepackage{amssymb}
\usepackage{array}
\usepackage{tikz}

\usepackage{bm}
\usepackage[colorlinks=true, linkcolor=blue, citecolor=green, urlcolor=blue]{hyperref}

\newcommand{\beq}{\begin{equation}}
	\newcommand{\eeq}{\end{equation}}
\newcommand{\bfig}{\begin{figure}[htbp]}
	\newcommand{\efig}{\end{figure}}

\newcommand{\ben}{\begin{eqnarray}}
	\newcommand{\een}{\end{eqnarray}}
\usepackage{color}
\usepackage{setspace}
\usepackage{comment}

\newcommand{\re}[1]{{\color{red}#1}}

\newif\ifincludecancelled
\includecancelledfalse 
\newcommand{\cancellable}[1]{\ifincludecancelled\sout{#1}\fi}

\usepackage{epsfig}
\usepackage[titletoc,toc,title]{appendix}
\usepackage{multicol}
\usepackage{mathtools}
\usepackage{environ}
\newtagform{Alph}[]()
\newtagform{roman}[]()
\newtagform{scroman}[][]

\begin{document}

	\title{\vspace{-4.5cm} }
\title{Global electromagnetic gyrokinetic simulations of internal transport barriers in reversed-shear tokamaks}

\author{Giovanni Di Giannatale\textsuperscript{1}, Arnas Vol\v{c}okas\textsuperscript{1}, Justin Ball\textsuperscript{1}, Alberto Bottino\textsuperscript{2}, Stephan Brunner\textsuperscript{1}, Philippe Griveaux\textsuperscript{1}, Moahan Murugappan\textsuperscript{1},   Thomas Hayward-Schneider\textsuperscript{2},  Laurent Villard\textsuperscript{1}}

\address{\textsuperscript{1} \`Ecole Polytechnique Féedérale de Lausanne (EPFL), Swiss Plasma Center (SPC), CH-1015 Lausanne,	Switzerland}
\address{\textsuperscript{2} Max-Planck-Institut für Plasmaphysik, D-85748 Garching, Germany}

    \ead{giovanni.digiannatale@epfl.ch}

\begin{abstract}
Internal transport barriers (ITB) {form} through turbulence suppression, {often observed when the}  safety
factor profile exhibits an off-axis minimum.
This work  {aims at improving our understanding of the conditions enabling the development of an ITB, using } a more comprehensive physical model, including low-$\beta$ electromagnetic flux-driven simulations. Our key findings are that electron dynamics is crucial {for ITB formation} even in an {ITG} scenario and that having $q_{\text{min}}$ {close to a {lowest order} rational value  (2 in our simulations) to allow for eddies self-interaction is a necessary ingredient.

Electron dynamics has two critical effects. {First, it leads to a structure formation characterized by strong zonal flows shearing rate, quench of turbulence (i.e. reduction of transport coefficients and fluctuation correlation) and profile corrugation. Second, it leads to zonal current sheets that result in}
 a broadening of the minimum-q region, qualitatively 
consistent with the flux-tube simulations of Volčokas et al. \cite{Volcokas_Arxive2024_magn}.}

Flux-driven simulations performed with $q_{\text{min}}=2$ reveal the development of the transport barrier in the ion channel, forming at inner and outer radial positions with respect to the $q_{\text{min}}$ position. The ITB formation in flux-driven setup is not recovered if $q_{\text{min}} = 2.03$. Additionally, a simulation at higher $\rho^*$ indicates that {the extent of the flattened region of the q-profile }
due to turbulent self-interaction does not change proportionally to $\rho^*$ or to $\rho_i$, but somewhere in between. On the other hand, the input power required to achieve similar on-axis temperatures appears to exhibit almost GyroBohm scaling (for the two considered $\rho^*$ values). 
 Furthermore, considering an initial q-profile with $q_{\text{min}} = 2.01$, flux-driven simulations show that partial self-interaction can evolve to complete self-interaction. This occurs due to  turbulent-driven zonal currents that lower and flatten the q-profile down to $q_{\text{min}} = 2.0$, in line with what is reported in Volčokas et al. \cite{Volcokas_Arxive2024_magn}.


\end{abstract}

\section{Introduction}
The safety factor is a critical parameter that affects confinement properties of fusion plasmas.
Since the first discovery in JET \cite{Hugon_nf1992}, a safety factor profile with a reversed shear region has been used to generate Internal Transport Barriers (ITB) in several devices. {In particular for JET reversed magnetic shear scenarios, ITB emergence occurs preferentially when the minimum $q$ reaches an integer value \cite{Joffrin_2003_nf}}.
These barriers have been experimentally observed both in the ion \cite{Koide_prl1994,Levinton_prl1995,Wolf_ppcf1999} and electron \cite{Hoang_prl2000,Litaudon_ppcf1996} heat channels.

 {Experimental observations have triggered} many theoretical and numerical works on the subject. It turned out that one of the basic stabilizing mechanisms for edge transport barrier formation relies on the decrease of the interchange drive with magnetic shear \cite{Drake_prl1996}. {Although the authors in  \cite{Drake_prl1996} focus on edge transport barriers, some commonalities with ITBs emerge. These include the local reversal of magnetic shear, which is driven by sufficiently large pressure gradients at the plasma edge}.

Analyzing the radial structure of the ITG mode, the authors in \cite{Romanelli_pop1993} already pointed out that the radial correlation length (and ultimately the transport coefficients) is strongly dependent on the magnetic shear and is reduced in the low shear case. This is the result of the decrease in the toroidal coupling between  poloidal harmonics. This explanation has been further expanded in \cite{Garbet_pop2001}, where the authors studied the formation of ITBs, focusing on the increase in the distance between resonant surfaces when the magnetic shear approaches zero. Indeed, for a given toroidal wave number $n$, the distance between two adjacent resonant surfaces is equal to $1/nq'$ {(valid at leading order but requiring second derivative effects for a zero-shear flux surface)}.

The hypothesis of resonant surfaces rarefaction has been disproved in \cite{Candy_pop2004}, where the authors investigated this hypothesis employing nonlinear gyrokinetic simulations with an adiabatic electron response. The authors demonstrated that, unless artificially ignoring non-resonant modes as done in \cite{Garbet_pop2001}, no such gap in rational surfaces exists, and that energy transport is smooth and increasing across such a minimum-$q$ region.

Subsequent simulations \cite{Waltz_pop2006}, performed with kinetic electrons, proved that what actually matters is the corrugation of the profiles. In the standard ITG adiabatic electron mode simulations, only weak corrugations in the ion temperature gradient are observed. They are much more pronounced when electron physics is included, because electron transport is more tightly coupled to the surfaces; the non-adiabatic response of passing electrons is strongest in the vicinity of low-order mode-rational surfaces and results in the emergence of zonal structures there \cite{Dominski_2017_pop}.

Around the rational surface  $q_\text{min}=2$ , there is a large profile corrugation and the buildup of a strong zonal ${\bf E}\times{\bf B}$ shear layer that could (at least partially) explain the strong reduction of the fluxes. The emergence of this localized poloidal flow is somehow expected within an ITB due to the link between the radial field and the ion pressure gradient arising from the force balance and it is now well  establish that $\bf{E \times B}$ shearing rate is one of the most effective phenomena  to quench turbulence \cite{Diamond_2005}.

Recently, the mechanism for the  onset of ITBs has been also studied with flux tube simulations \cite{Volcokas_nf2023,Volcokas_Arxive2024}. The authors studied the importance of  {correctly modeling field line topology} when considering rational values of the safety factor in order to have the correct eddies self-interaction that ultimately leads to a dramatic change of the heat flux. 
A flux-tube study of the sensitivity of the safety factor around rational values has been addressed in \cite{Volcokas_Arxive2024} and it is shown that  going slightly above $q=2$ can produce twice the fluxes in some circumstance (e.g. CBC parameters) or it can reduce fluxes in other cases (e.g.  pure ITG {-meaning only ion temperature gradients being different than zero-}). While we recover the sensitivity to the rational values of $q$, we actually find the {ITG} case {(both ion and electron temperature gradients different than zero)} being stabilized when operating at $q_\text{min}=2$ compared to the case $q_\text{min}=2.03$.

Most of the modeling so far employed to study conditions relevant to the onset of ITBs has involved electrostatic perturbations {and works including non-zero $\beta$ (e.g. \cite{Waltz_pop2006}) do not focus on electromagnetic effects}. The first simulations {that  also include a detailed analysis of some electromagnetic effects}  are presented in \cite{Volcokas_Arxive2024_magn} and in this paper. As shown in \cite{Volcokas_Arxive2024_magn}, flux-tube small {magnetic} shear turbulence saturation is strongly connected to the generation of the parallel current that, in turn, modifies {locally the safety factor profile}, leading to an {additional} contribution to turbulence reduction even at small $\beta$ ({where electrostatic turbulence dominates}). However, global simulations presented in this work do not seem to be particularly affected by the flattening of the $q$-profile as long as both the nominal and the modified $q$-profile allow for a {complete} self-interaction. {On the other hand, they are  affected for cases where the modified $q$-profile allows for a transition from partial self-interaction to full self-interaction.}

With this work, we aim to perform a systematic study concerning the effects of a reversed safety factor profile, relevant to ITB formation. We will employ three different models for the electron response: fully adiabatic, hybrid, and fully drift-kinetic. The models are detailed in Section 2. 

With the {adiabatic electrons} setup, we can reasonably quickly scan several parameters concerning the safety factor, such as the impact of the  $q_\text{min}$  values and the curvature radius around the  $q_\text{min}$ ({although not reported in this paper since with adiabatic electrons we found practically no difference when changing the curvature of around $q_\text{min}$}). Global effects will be emphasized, especially with the hybrid and the fully kinetic models, where a small change in the safety factor around the minimum position has an enormous effect on the whole plasma domain, varying the turbulent transport by a factor of about two. This feature is observed with both the hybrid and the fully kinetic models.

While in the first part of this work we present gradient-driven simulation, the emphasis of this paper is in the last section, where we  present flux-driven simulations successfully reproducing the ITB establishment.
A $\rho^*$ comparison (DIII-D versus TCV) is also  analyzed. {The same electromagnetic effects that lead to $q$-flattening around $q_\text{min}$, as observed in flux-tube simulations \cite{Volcokas_Arxive2024_magn}, will be described.}

This contribution is structured as follows. In the next section, we provide details on the global gyrokinetic code ORB5 and the simulation setup. In Section \ref{adiabatic_section_results}, the adiabatic electron simulations are presented. In Section \ref{hybrid_section_results}, we describe the results obtained using the hybrid electron model. Here, we shall see the effect of having  $q_\text{min}=2$  and its global impact. Fully kinetic simulations are reported in Section \ref{fullykin_section_results}. Here, we will perform the comparison between electrostatic and electromagnetic simulations. Flux-driven simulations and the ITB formation are described in Section \ref{fluxdriven_section_results}, where we compare  simulations with  and without ITB. An additional simulation using TCV-like $\rho^*$ is also shown. \cancellable{Finally, in the flux-driven section we also present the effects of electromagnetic perturbations when $q_\text{min}$ allows for a partial (but not complete) self-interaction.}

Finally, we carry out a flux-driven simulation starting from an initial state with $q_\text{min}=2.01$, which allows partial self-interaction, and show that the $q$-profile evolves towards $q_\text{min}=2.0$, allowing for complete self-interaction to take place.
Conclusions are drawn in Section \ref{section_conclusions}.

\section{Numerical setup and case description} \label{section_setup}
The global gyrokinetic simulations presented in this work  {were} performed with the ORB5 code \cite{ORB5}. ORB5 is a global gyrokinetic code that uses a PIC approach and finite element representation.
It solves the full-$f$ Vlasov equation in spite of the $\delta f $ splitting {of $f$ into $f_0 + \delta f$, with $f_0$} used as control variates; while the polarization term of the quasi-neutrality equation  is linearized around the axisymmetric part of the distribution function $f(r,t)$, that evolves in time.

The finite elements linear systems of equations, {resulting from the discretization of the quasi-neutrality and Amp\`ere equations}, are projected into {toroidal and poloidal} Fourier space in order to decouple the various {toroidal} harmonics and  save computational time by retaining only the modes of interest \cite{MCMillan_2010, ORB5}. 
The electromagnetic perturbations are efficiently handled using the pullback scheme \cite{Mishchenko_2019}.

The magnetic equilibrium in ORB5 is defined as follows:
\begin{equation}
    {\bf B} = F(\psi) \nabla\phi + \nabla \psi \times \nabla \phi
\end{equation}\label{equil_adhoc}
where $F(\psi)$ is the poloidal current flux function, $\psi$ is the poloidal
magnetic flux and $\phi$ is the toroidal angle. The ORB5 code uses
ideal-MHD equilibria, solutions of the Grad–Shafranov equation,
which are provided by the CHEASE code \cite{Lutjens_1996}. It can also use
an analytical ad-hoc magnetic equilibrium comprising circular, concentric magnetic surfaces. A straight-field-line coordinate system is used in ORB5. The
magnetic surfaces are labeled by $s=\sqrt{\psi/\psi_{\text{edge}} }$, where $\psi_{\text{edge}}$ is
the value of $\psi$ at the radial edge, and the
straight-field-line poloidal angle is defined by
\begin{equation}
    \theta^* = \frac{1}{q(s)} \int_0^\theta  \frac{B^\phi}{B^\theta}  d\theta'  
\end{equation}

with $B^\alpha = {\bf B} \cdot \nabla \alpha$. In this work we use an ad-hoc geometrical equilibrium with a prescribed safety factor. 
It consists of circular, concentric magnetic surfaces, with $ {\text{d}\psi}/{\text{d}r} = {r B_0}/{\bar{q}(r)} $, where $r$ is the minor radius and $\bar{q}(r)$ is a user specified function. The safety factor ${q}(r)$ is then internally computed and it differs from $\bar{q}(r)$ by second order terms in $r/R$ (with $R$  the major radius):

\begin{equation}
    q(s) = \frac{\bar{q}(r)}{\sqrt{1- (r/R)^2}}  \;.
\end{equation}

The safety factor $\bar{q}(r)$ we choose is a $5^{th}$ order polynomial. The coefficients of the polynomial are chosen to target specifically $dq/ds = 0$ at $r=0$ and at $  r=r_{\text{min}}$, a given value of $q$ on axis, $q(0)$, and to have a certain freedom on the choice of the $q$-curvature (namely $d^2q/ds^2$) around  $r_{min}$ and on the possibility to keep the magnetic shear $r/q \;dq/ds$ unchanged when changing the $q_{\text{min}}$. Here $r_{\text{min}}$ indicates the position of the minimum of $q$ and $q_{\text{min}} := q(r_{\text{min}})$. 

In particular, we prescribe a safety factor that has a minimum, and thus zero shear, at $s=0.52$. We will perform a scan in the $q_{\text{min}}$ value (but keeping the same shear profile) or changing the extent of the weak shear region (namely acting on safety factor second derivative). We focus our study on cases  when $q_{\text{min}}$ corresponds to a low order rational value, namely $q_{\text{min}}=2$,  or values just above 2.

\begin{figure}[htbp]
    \centering
    \begin{adjustbox}{addcode={}{},left}
        \hspace{-0cm}
        \includegraphics[width=1.\textwidth]{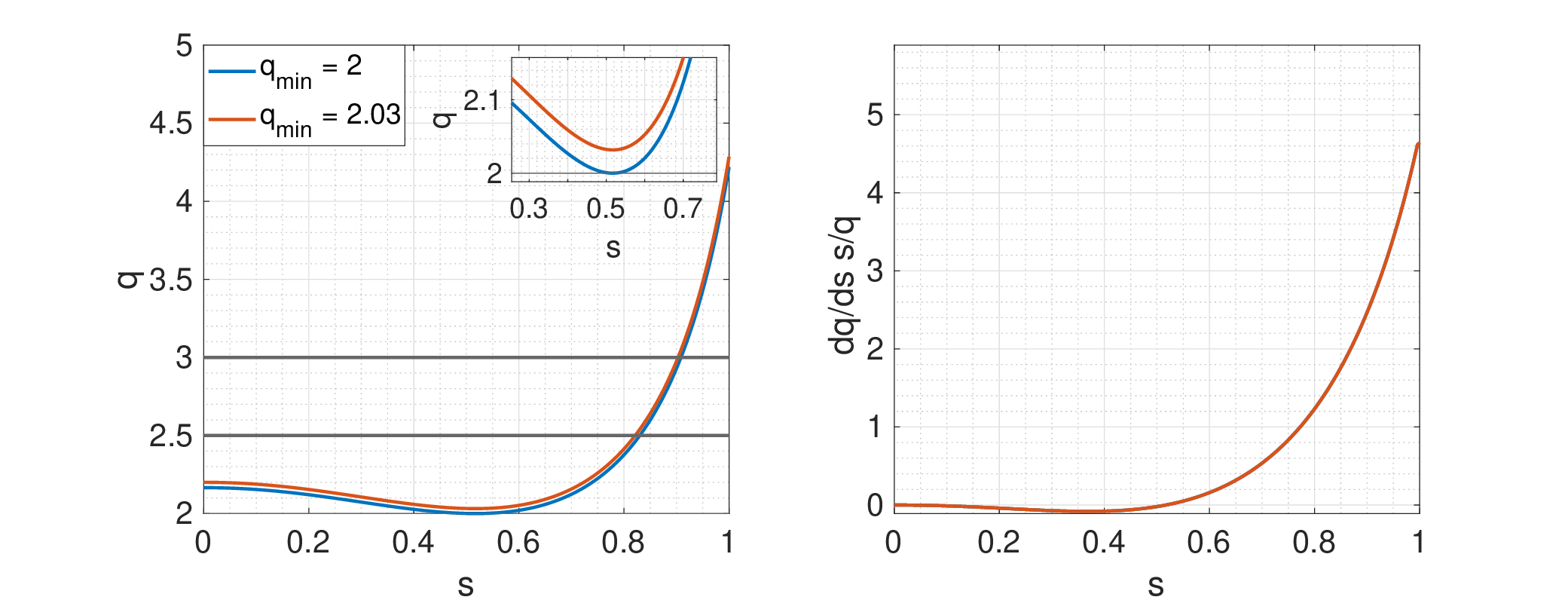}
        \begin{tikzpicture}[overlay,remember picture]
        \end{tikzpicture}
        
    \end{adjustbox}
    \caption{{Safety factor (left) and magnetic shear (right) profiles for two different $q_{\text{min}}$, $q_{\text{min}} = 2.00$ and $q_{\text{min}} = 2.03$ , blue and red respectively. A zoom of the safety factor profiles around the $q_{\text{min}}$ region is also shown.}}
    \label{linear}
\end{figure}

{The temperature profiles are obtained integrating the prescribed logarithmic gradients $R/L_T$, that have a functional form defined as a function of $r=\rho_{vol}$, ({defined as $\rho_\mathrm{vol}=\sqrt{V(\psi)/V(\psi_a)}$, where $V(\psi)$ is the volume enclosed by the magnetic surface $\psi=\mathrm{const}$.}):

\begin{equation} \label{functional_form}
    \frac{R}{aT} \frac{dT}{dr} = -\frac{\kappa_T}{2} \left[ \tanh{\left(\frac{r_{-}}{\Delta_T}\right)} - \tanh{\left(\frac{r_{+}}{\Delta_T}\right)}  \right],
\end{equation}
with $r_{\pm} = r -r_0 \pm \Delta_r /2$. }

The simulations parameters are inspired from a {zero density gradient}  Cyclone Base Case (CBC) scenario, i.e. $R/L_T = 6.9 $, $\rho^* = 1/185$, $R/L_n = 0$ with $L_x= -x/(\text{d}x/\text{d}r)$ {in the flat part of the logarithmic gradient profiles}. The gradients are the same for both ions and electrons. In addition, collisionless dynamic only is considered.

For this work we employed all  three models for the kinetic species that are available in ORB5, namely the adiabatic electron model , the hybrid electron model and the fully kinetic electron model. While in the adiabatic model  {all electrons give a Boltzmann } response  to the quasi-neutrality equation (QNE), the hybrid model works as follows: 

\begin{itemize}
    \item the FLR effects of electrons are neglected (electrons are drift-kinetic);
    \item all electrons are evolved according to the GK equations;
    \item in the QNE the non-zonal passing electrons contribution is assumed to be adiabatic, but the zonal contribution of all electrons (passing and trapped) is retained.
\end{itemize}
Finally, in the fully kinetic model the only assumption is that the electrons are treated drift-kinetically.

While in the second part of the paper we  focus on flux-driven simulations, we start with Temperature-gradient-driven runs. 
 For a global full-f code the concept of gradient-driven has to be taken carefully: one starts from a certain temperature profile and  a heating operator is then applied in order to maintain this profile. This operator has a Krook form: $S[\delta f, f_0]= - \gamma_K \,\delta f + S_{corr}[\delta f,f_0]$. The term $\gamma_K \delta f $ holds the temperature close to the initial one, while the operator $S_{corr}[\delta f,f_0]$ acts as a correction term to ensure that the whole operator $S[\delta f, f_0]$ does not affect {flux-surface-averages of} zonal flows, parallel momentum and density \cite{McMillan_2008}.
The coefficient $\gamma_K$ is set to less than $ 10\% $ of the maximum linear growth rate. Since with this operator a certain level of relaxation of the temperature profiles is allowed, the most relevant (and fair) quantity to be compared {between two simulations} {starting from the same temperature profile} is the heat diffusivity $\chi$ defined as an effective local heat diffusivity:

\begin{equation}
    \chi = - \frac{\langle Q_H \cdot \nabla \psi \rangle}{n \frac{dT}{d\psi} \langle \left|\nabla \psi \right|^2 \rangle} \; ,
     \label{chi_computation}
\end{equation}
with $Q_H$ standing for the heat flux {and $\langle \cdot \rangle$ for the flux surface average operator}. We point out that calling $\chi$ an {\itshape effective} diffusivity does not mean that transport is purely diffusive.  A more  {detailed} analysis is presented in the sections below. \\

As we shall see, fully kinetic simulations lead to a strong corrugation of temperature profile, especially when running in flux-driven and there is no constraint on temperature profiles. This may lead to a strong  {deviation of the temperature (and density, parallel flows, etc) from their initial profiles, and therefore large values of $|\delta f|$}. When these deviations become too big, then the advantage of the global $\delta f $ PIC approach can be lost. Indeed,  the gain in noise reduction of the $\delta f $ scheme relies on the reduced
variance of the marker weights, {which can occur only } provided that the assumption $||\delta f|| / ||f||<<1$ for some definition of the norm $||\cdot ||$ is met. In order to still possibly retain some advantage of the delta-f scheme, one could also evolve  $f_0$, albeit at a longer time scale than that of the fluctuating $\delta f $ \cite{Brunner_pop1999,Allfrey_cpc2003}. In the fully kinetic electron runs of this work, we exploit the benefits of having a time-evolving background by constraining $ f_0 $ to be a flux-surface-dependent Maxwellian which is time-dependent via its evolving gyrocenter density and temperature profiles. Details on the numerical aspects, findings and assumptions of the control variate adaptation scheme implementation in ORB5 are described in  Refs.    \cite{Murugappan_Arxiv2024,Murugappan_2022_pop}.



\section{Adiabatic electrons simulations} \label{adiabatic_section_results}

\begin{figure}[htbp]
\centering
    {\includegraphics[width=.5\textwidth]{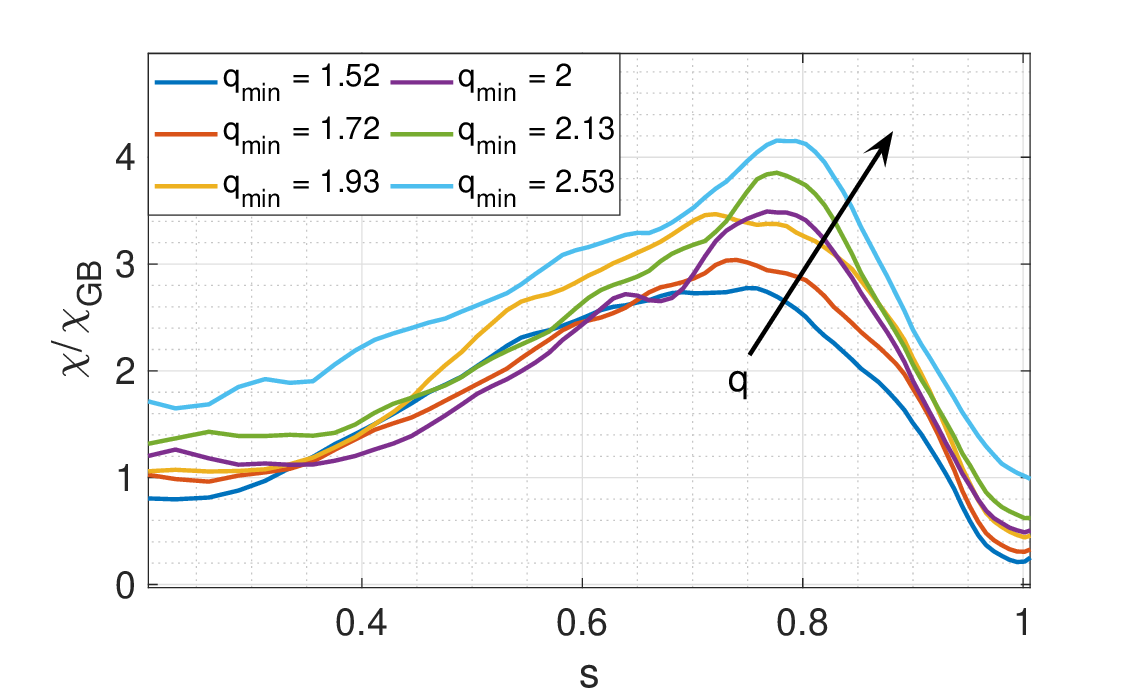}}
\caption{{Ion transport coefficient for simulations performed by varying the value of  $q_{\text{min}}$, as indicated in the legend. The q-profile is modified while keeping  the magnetic shear constant (as for all the other simulations). Gradient-driven simulations.}}
\label{non_linear}
\end{figure}

Our analysis starts by considering the adiabatic electrons response.
The first assessment concerns the sensitivity of the system with respect to the $q_{\text{min}}$ value. In this scan, $q_{\text{min}}$ has been changed (while keeping  the magnetic shear constant) to evaluate whether being a rational value (or not)  affects the transport coefficients and whether there is a strong discontinuity in the fluxes when varying $q_{\text{min}}$. The results, summarized in Figure \ref{non_linear}, show that with an adiabatic electron response \cancellable{the system does not particularly discriminate between different q profiles and} {having $q_{\text{min}}=2$ (i.e. on a rational surface) or not} does not lead to any particular difference. The situation will appear different in the next sections, where we include a non-adiabatic electron response.
However, a safety factor effect can still {be observed, as there is} a well-established trend in the transport coefficient $\chi$, which progressively increases with $q$ (see Figure \ref{non_linear}).


\section{Hybrid electron model simulations} \label{hybrid_section_results}
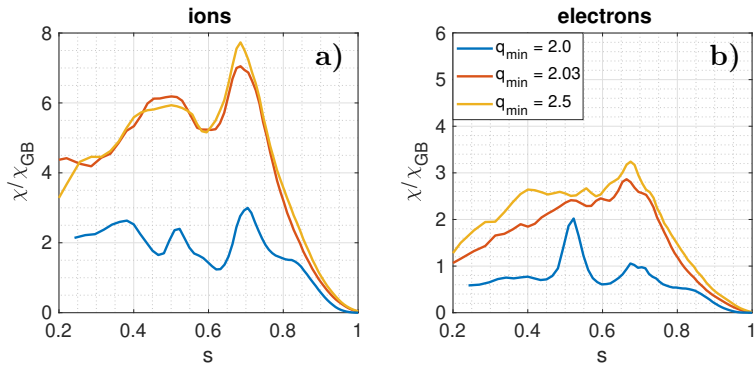
\begin{figure}[htbp]
    \centering
    \begin{adjustbox}{addcode={}{},left}
        \hspace{-0cm}
        \includegraphics[width=0.8\textwidth]{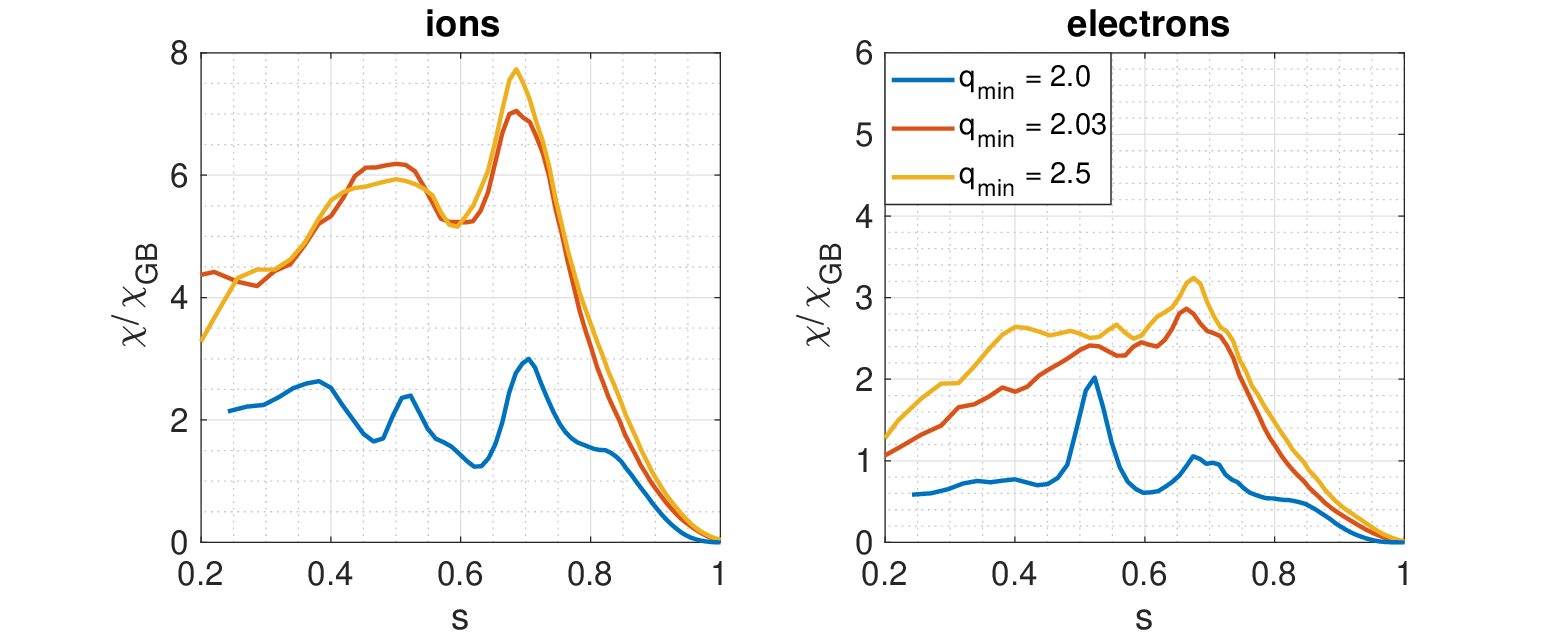}
        \begin{tikzpicture}[overlay,remember picture]
            \node at (-6.8, 4.1) {\textbf{a)}}; 
            \node at (-1.6, 4.1) {\textbf{b)}}; 
        \end{tikzpicture}
        
    \end{adjustbox}
    \caption{Ion and electron transport coefficients for three different q-profiles, with $q_{\text{min}} = 2 , \;2.03, \;2.54$ (blue, red and yellow, respectively). {Contribution to $\chi$ from ${\bf E} \times {\bf B}$ fluctuations only. Gradient-driven simulations.}}
    \label{q_scan_hybrid}
\end{figure}

We focus on the effects of different values of  $q_{\text{min}}$ while keeping the magnetic shear constant, as in Figure \ref{non_linear}. We examine three $q_{\text{min}}$ values: $q_{\text{min}} = 2,\; 2.03,\; 2.54$. The results are shown in Figure \ref{q_scan_hybrid}. It is evident that using $q_{\text{min}}=2$ can lead to a significant reduction in the transport coefficients for both the ion and electron channels. The global effect of  $q_{min}=2 $ is also notable: the transport {reduction} is achieved across the entire plasma radius. Although this effect is particularly surprising, we stress that a very similar result has already been  obtained in  Ref. \cite{Waltz_pop2006}.

Remarkably, with  $q_{\text{min}}=2.03$ {no significant differences are observed compared to the $q_{\text{min}}=2.54$ case}. This is due to the fact that self-interaction of eddies is lost. Since eddies have a final size in binormal and radial directions, the self-interaction can persist even if $q_{\text{min}}$ is not exactly 2, and the maximum value of  $q_{\text{min}}$ will  depend on the eddy size and thus on $\rho^*$.

Analyzing the {evolution in space and time} of the two simulations with slightly different $q$ (namely the cases  $q_{\text{min}} = 2,\; 2.03$), one can notice that the transport reduction starts to occur when the linear phase saturates in the region near the zero shear position. The saturation leads to a strong, steady $v_{{ E} \times { B}}$ shearing rate $\omega_{{ E} \times { B}}$ surrounding the zero shear location, with

\begin{equation}    
 \omega_{{ E} \times { B}} = \frac{s}{2 \psi_{\text{edge}}\, q(s)} \frac{\partial}{\partial s} \left (\frac{1}{s}\frac{\partial \langle \phi \rangle}{\partial s} \right ) .
\end{equation}

\begin{figure}[htbp]
    \centering
    \begin{adjustbox}{addcode={}{},left}
        \hspace{2cm}
        \includegraphics[width=0.7\textwidth]{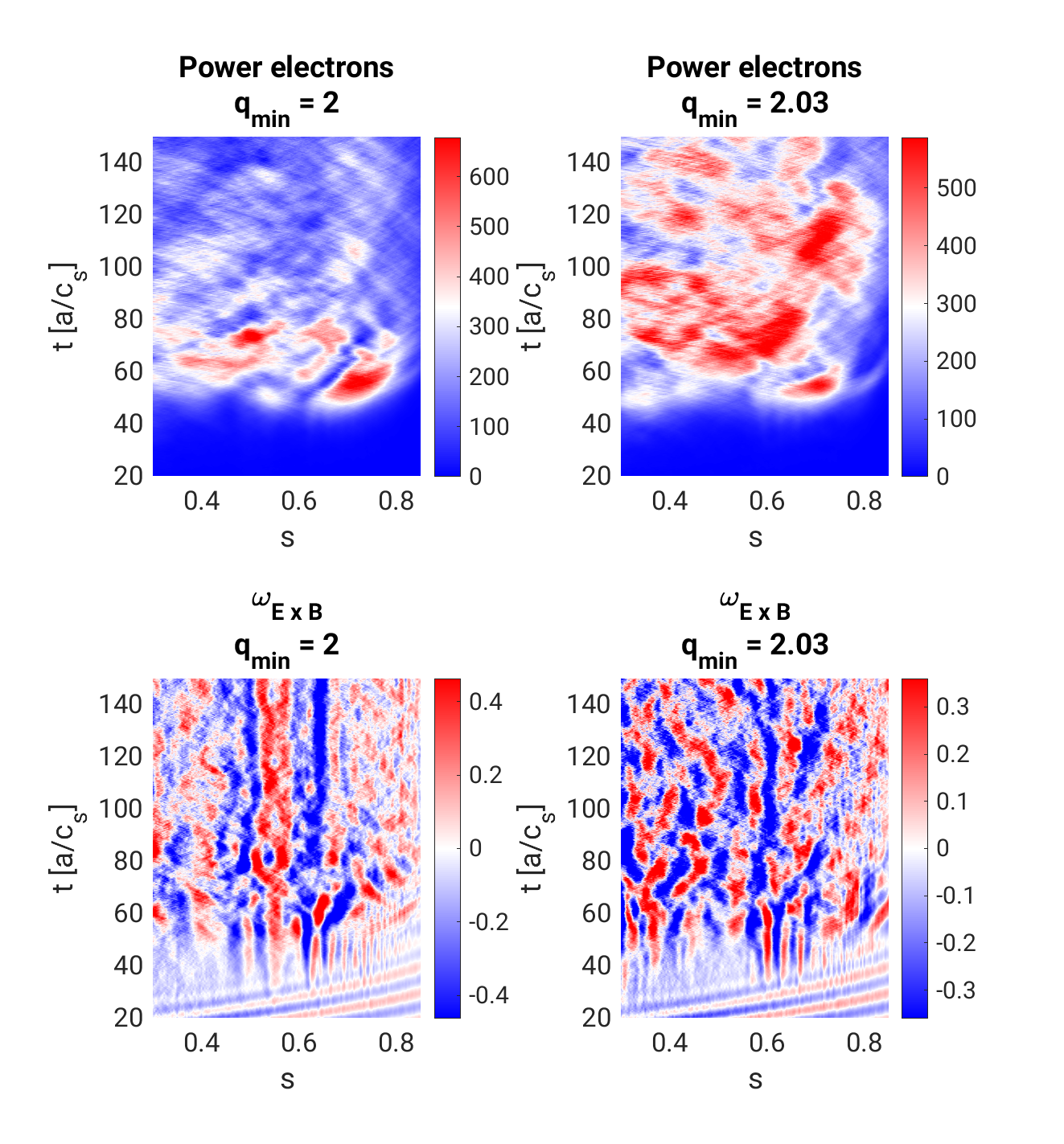}
        \begin{tikzpicture}[overlay,remember picture]
             \node at (-1.5, 10) {\textbf{b)}}; 
             \node at (-6.0, 10) {\textbf{a)}}; 
             \node at (-6, 4.8) {\textbf{c)}}; 
             \node at (-1.5, 4.8) {\textbf{d)}}; 
        \end{tikzpicture}
        
    \end{adjustbox}
    \caption{{2D (time and radial coordinates) plots of the heat powers and  $E \times B$ shearing rates ($\omega_{E \times B})$, upper and lower panels, respectively. Left $q_{\text{min}} =2 $, right $q_{\text{min}} =2.03 $. Gradient-driven simulations. }}
    \label{2D_power_and_ZF}
\end{figure}
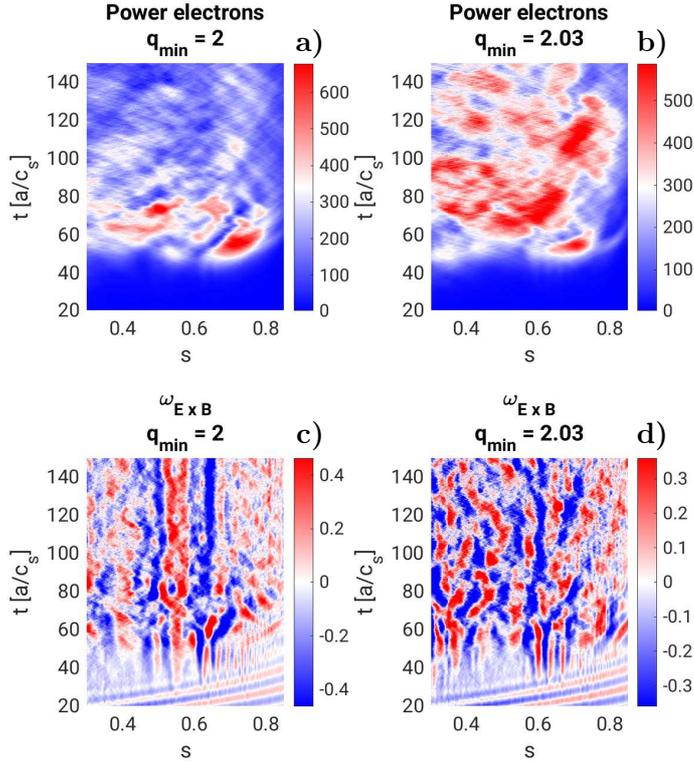

This is shown in Figure \ref{2D_power_and_ZF}. Around $t \,c_s/a = 70$, the turbulence saturates and a large $\omega_{E \times B}$ dipole  develops around $q_{\text{min}}$, with $\omega_{E \times B} = 0 $ at $s_{\text{min}}$. Comparing the case $q_{\text{min}} = 2$ (Figures \ref{2D_power_and_ZF}-a,c) and the case $q_{\text{min}} = 2.03$ (Figures \ref{2D_power_and_ZF}-b,d), 
it is clear {that} there is a difference in  $\omega_{{ E} \times { B}}$ around $s_{\text{min}}$ (0.52)  and this difference  {is reflected} in the fluxes (panels a,b). In the case of $q_{\text{min}} = 2.03$ (panel b), no significant changes occur, whereas in the case of $q_{\text{min}} = 2$ (panel a), there is a substantial drop right after $t \, c_s/a = 80$, following the creation of the large, steady $\omega_{{ E} \times { B}}$.

\begin{figure}[htbp]
\centering
\begin{tikzpicture}
    \node[inner sep=0pt] (image1) at (0,0)
    {\includegraphics[width=1\textwidth]{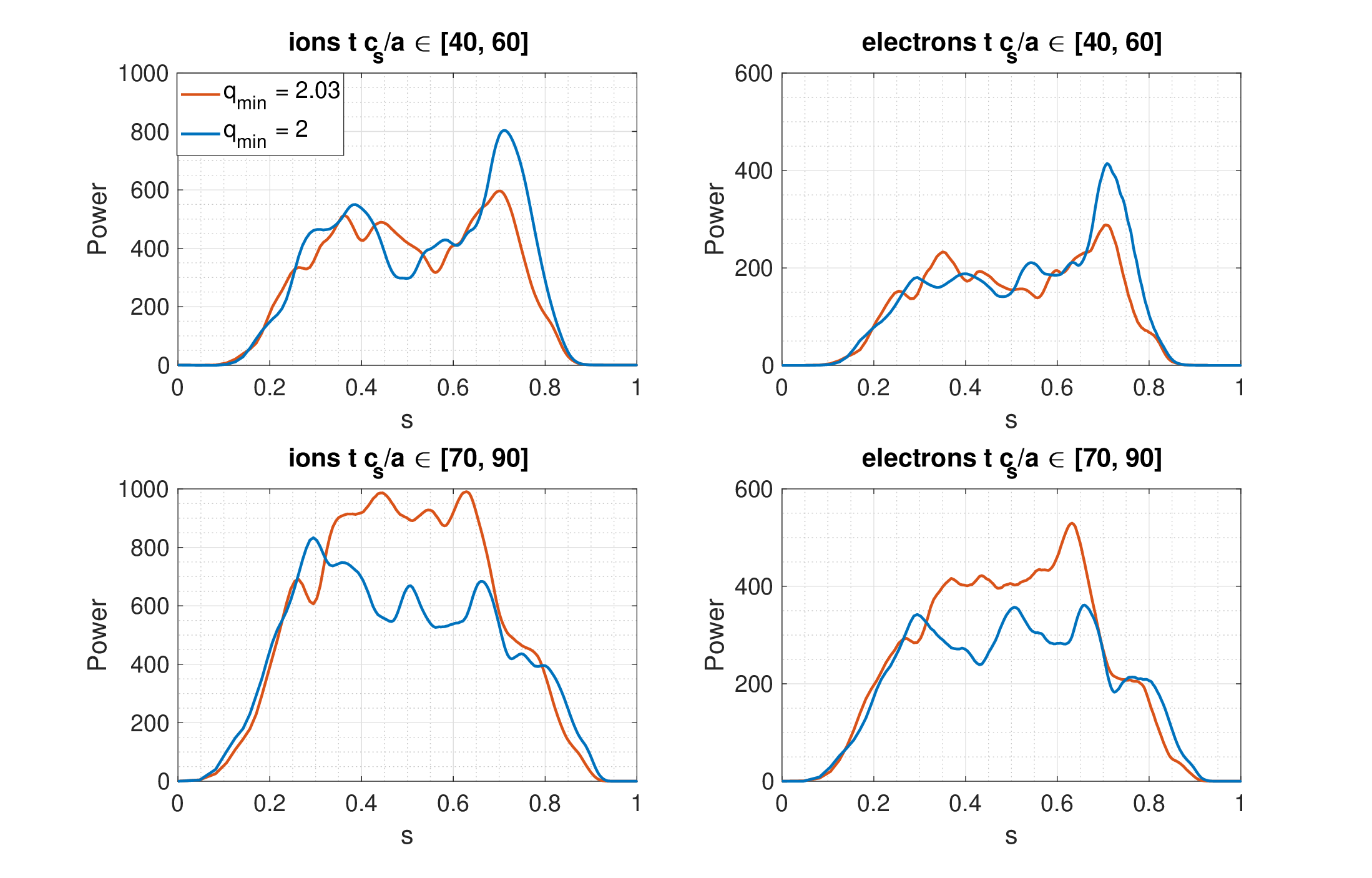}};
             \node at (5.5, 3.6) {\textbf{b)}}; 
             \node at (-1.0, 3.6) {\textbf{a)}}; 
             \node at (-1, -0.85) {\textbf{c)}}; 
             \node at (5.5, -0.85) {\textbf{d)}}; 
\end{tikzpicture}
\caption{{Heat flux  for ions (left) and electrons (right) for $q_{\text{min}} =2$ (blue) and   $q_{\text{min}} =2.03$ (red). Two different time windows are shown $t\, c_s/a \in [40-60]$ (top panel) and $t\, c_s/a \in [70-90]$ (bottom panel). Gradient-driven simulations.}}
\label{power_different_times}
\end{figure}

{Looking at the power through flux surfaces at different time intervals as shown in Figure \ref{power_different_times}, one can see the role of such $\omega_{E \times B}$ dipole. On the top panels an early time {interval is shown, $t \,c_s/a$ averaged $\in[40,60]$, during which}   there is not a significant difference between the two configurations; on the contrary, the bottom panels refer to an average time centered at the moment of the  ${\bf E} \times {\bf B}$ saturation, namely $t\, c_s/a \in [70, 90]$}. {This again demonstrates the non-linear global transport reduction and stabilizing effect of $q_{\text{min}} $.}

\section{All electrons drift-kinetic  simulations} \label{fullykin_section_results}
{Even though the hybrid model does include partial effects associated with kinetic electrons, it neglects the kinetic response of passing electrons in the  QNE for the non-axisymmetric modes.}
This can lead to inaccurate estimations of features that are strongly connected with non-adiabaticity of electrons. 
In our case, including a full kinetic response might be necessary since the adiabatic condition,  $|\omega/k_\parallel| \ll v_{th,e}$, is clearly violated around a {mode} rational surface 
{where $|k_\parallel| \rightarrow 0$. For shear-reversed profiles, the region where the adiabatic condition is violated is large if $q_{min}$ is near a low-order mode rational surface.}
With the fully kinetic electron model we  perform both gradient-driven and flux-driven simulations (described in the next section) using the adaptation scheme briefly described in section \ref{section_setup} (more information in \cite{Murugappan_Arxiv2024,Murugappan_2022_pop}). 

The main difference with respect to the hybrid simulations is the inclusion of the electromagnetic effects that develop even at small $\beta$.
In particular, it is interesting to see how including electromagnetic perturbations leads to a zonal, steady magnetic field that modifies the effective safety factor profile seen by particles. {As we will show, this modified $q$-profile is a result of turbulence-generated currents.} Indeed, turbulence creates a strong electron $v_\parallel$ dipole, that translates to currents that in turn result in a perturbed magnetic field thanks to Amp\`ere law. {Further details of the mechanism generating turbulent currents are discussed in Ref. \cite{Volcokas_Arxive2024_magn}.}

The modified safety factor can be computed either through the equations of motion (as done in \cite{Ball_ppcf2023}) or using the classic definition:
\begin{equation}
    q = \frac{\bf B  \cdot \nabla \phi}{ \bf B \cdot \nabla \theta^*} \quad.
\end{equation} \label{safety}
{Separating the equilibrium (0) and perturbed (1) contributions, we get}
\begin{equation}
q = \frac{B^\phi_0 + B^\phi_1}{B^\theta_0 + B^\theta_1} = \frac{B_0^\phi B_0^\theta + B^\phi_1 B_0^\theta +  B_1^\theta B_0^\phi +  B_1^\phi B_1^\theta}{(B_0^\theta)^2 - ( B_1^\theta)^2} \simeq q \,( 1 - \frac{B_1^\theta}{B_0^\theta}) + \frac{B_1^\phi }{B_0^\theta} 
\end{equation} \label{safety_2}
where in the last equality quadratic terms, $B_1^j B_1^k$, have been neglected coherently with the {gyrokinetic ordering used}.  The terms $B_1^j $ are computed considering only steady zonal $A_\parallel(\psi)$.  

In this section we focus on gradient-driven results. Both purely electrostatic and small $\beta $ electromagnetic simulations have been performed. These simulations have the same physical setup as described in the hybrid model. Furthermore,  here the neoclassical term has not been considered; namely the following assumption is made: { $(\text{d}/\text{d}t)|_0 \,f_0 =0 $, where the derivative is taken along unperturbed orbits}. {In  \ref{A_1} we show an additional simulation including the $(\text{d}/\text{d}t)|_0 \,f_0  $ term, and we briefly describe how it affects the system. For the reader who wants to skip the section, we found that including the term does not affect the validity of the other simulations and it does not change the main conclusions of our work.}

\subsection{Electrostatic}
In this section we compare the hybrid and the electrostatic (ES) fully kinetic electron models {(with $\beta=0$)} for the $q_{\text{min}}=2$ case. Compared to the hybrid model, the fully kinetic electron model leads to higher fluxes and transport coefficients. This effect is further increased since we use heavy electrons ($m_i/m_e = 500$). While a systematic evaluation of the heat flux scaling versus electron mass is difficult, it has been observed that heavy electrons  lead to an overestimate of the growth rate. For the real mass ratio and away from rational surfaces, the ITG and TEM {(although they are linearly stable in this work)} growth rates  converge towards the hybrid model response \cite{Bottino_ppcf2011}. According to our experience, this reflects also on the turbulent transport coefficients.

\begin{figure}[htbp]
    \centering
        \includegraphics[width=1\textwidth]{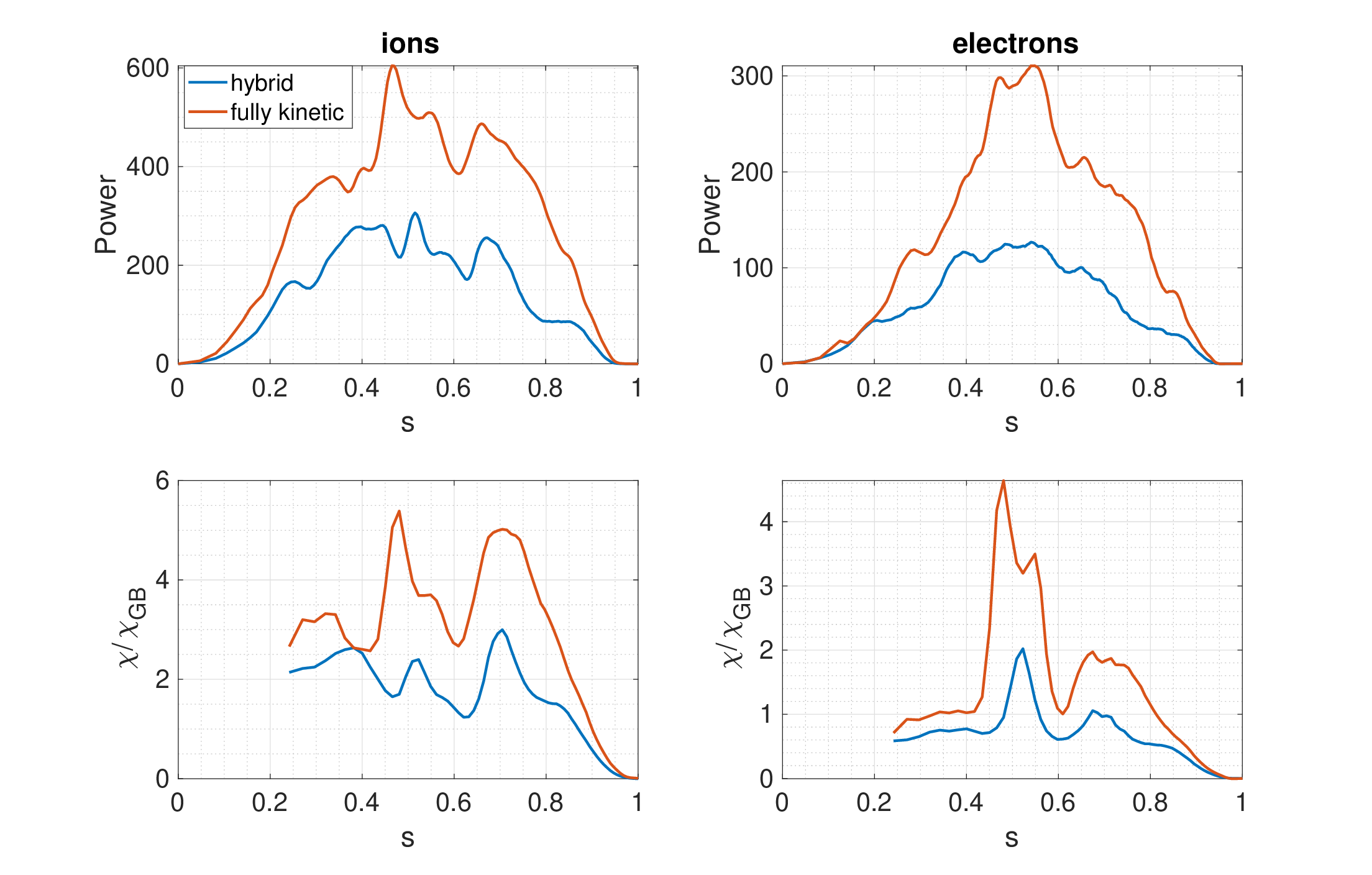}
        \begin{tikzpicture}[overlay,remember picture]

        \end{tikzpicture}
        
    \caption{Comparison of heat power (top) and transport coefficients (bottom) between the fully kinetic and hybrid models for ions (left) and electrons (right).  The $q_{\text{min}}=2$ case is considered. Gradient-driven ES simulations.}
    \label{hybrid_fullyk}
\end{figure}

\begin{figure}[htbp]
    \centering
        \includegraphics[width=0.8\textwidth]{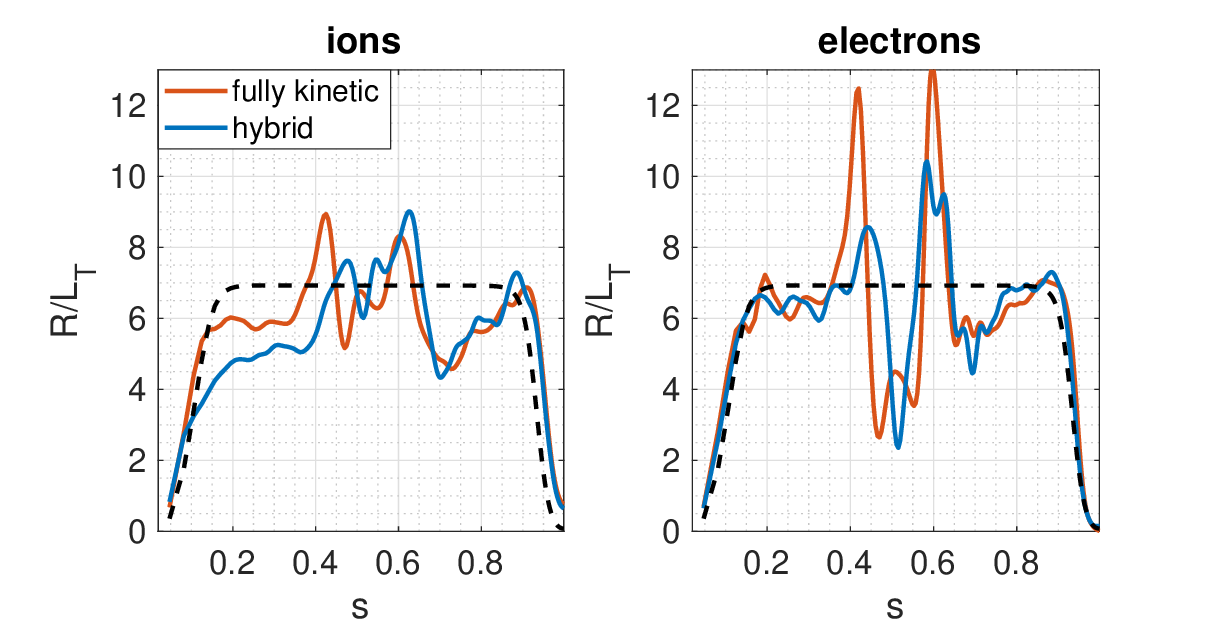}
        \begin{tikzpicture}[overlay,remember picture]
        \end{tikzpicture}
        
    \caption{{Comparison of logarithmic gradients between the hybrid and fully kinetic models (same simulations as shown in Figure \ref{hybrid_fullyk}). Left: ions; right: electrons. The dashed line represents the initial logarithmic gradients. Gradient-driven ES simulations.}}
    \label{temperature_gradients}
\end{figure}

\begin{figure}[htbp]
    \centering
        \includegraphics[width=0.5\textwidth]{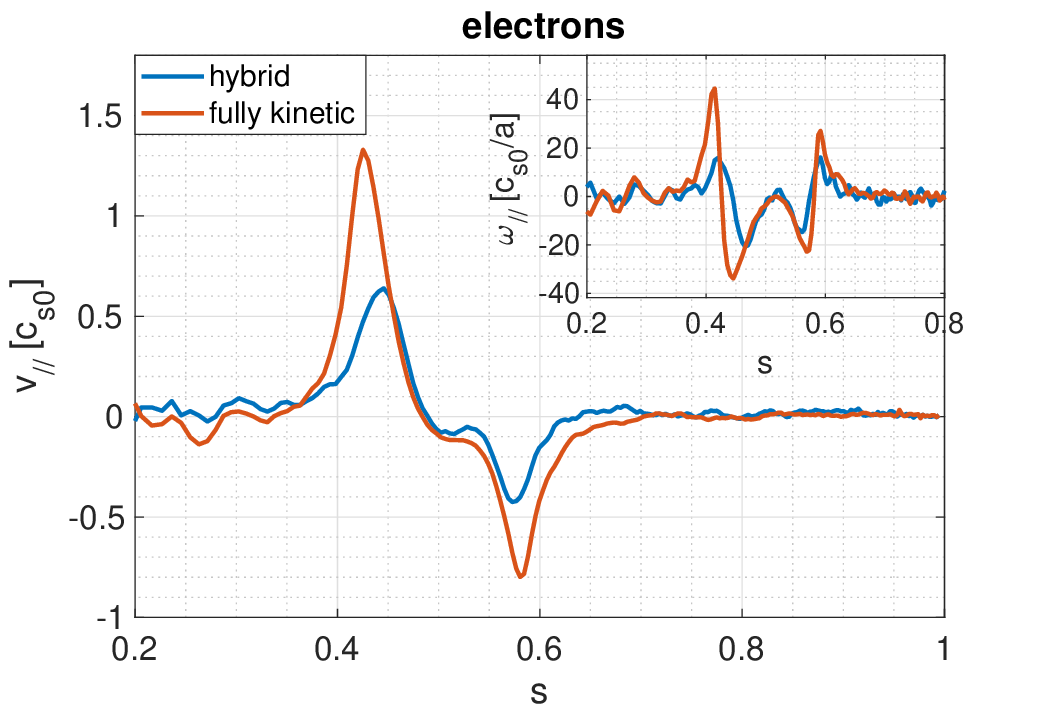}
        \begin{tikzpicture}[overlay,remember picture]
        \end{tikzpicture}
        
    \caption{{Comparison of the electron parallel velocity between fully kinetic and hybrid models. In the subplot the radial derivative $\text{d}v_\parallel/\text{d}s $ is shown. Gradient-driven ES simulations. }}
    \label{v_parallel}
\end{figure}

However, a comparison with the hybrid model can still be done and it shows a qualitative  agreement. The heat power and the transport coefficients are shown in Figure \ref{hybrid_fullyk}. Even though the hybrid model underestimates  the  fluxes and the transport coefficients  {by a factor of 2}, a remarkable qualitative agreement emerges from the comparison. Profile corrugations and transport barriers are found in both models; {however, they are significantly} {more pronounced} for the fully kinetic case, as shown in Figures \ref{temperature_gradients}-\ref{v_parallel}. This {indicates that an} \cancellable{is strong indication that the} important part of the {non-adiabatic} electron response (for the ITB onset) is related to the zonal component {of the electrostatic field}. Indeed, in the hybrid model that we employ the zonal potential is computed using a fully kinetic response (the adiabatic approximation for the passing electrons is used for the $n \not = 0$ modes).

For the fully kinetic run, the radial drop in the transport coefficients (see equation \ref{chi_computation}) for the electron channel is impressive: {$ \chi$ decreases } from $ \chi \simeq 3.3 \, \chi_{\text{GB}} $ near $s=0.56$ to $ \chi \simeq 1 \, \chi_{\text{GB}} $ near $s=0.6$, i.e. a change of $ \Delta \chi  \simeq 2.3 \, \chi_{\text{GB}}$ within  only $\simeq 9 \rho_i$.

\begin{figure}[htbp]
    \centering
        \includegraphics[width=1\textwidth]{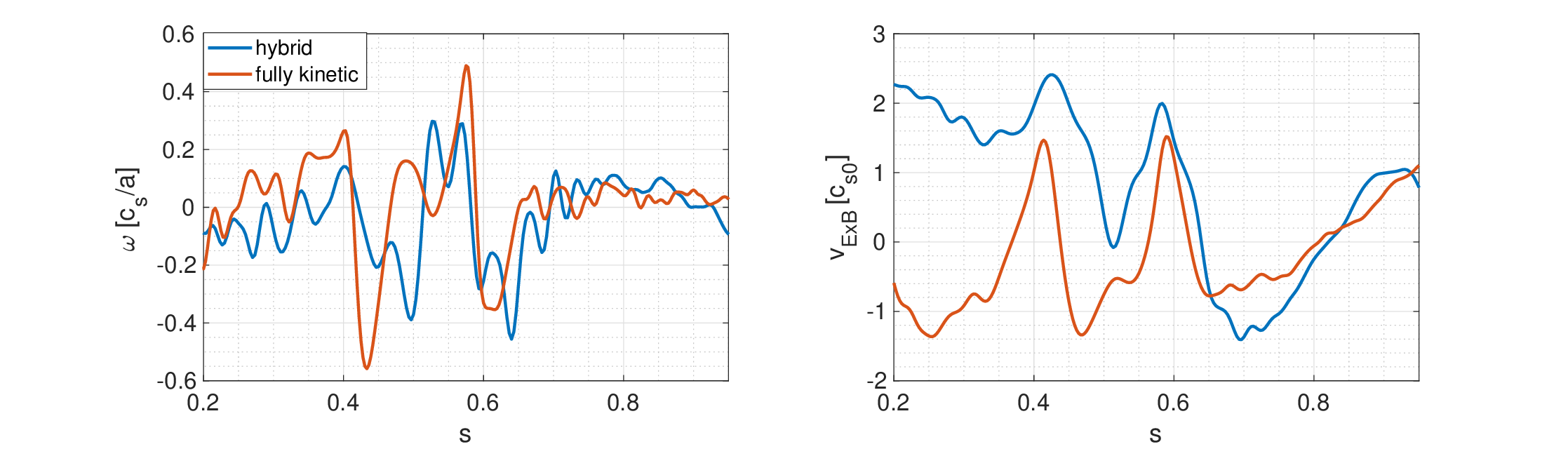}
        \begin{tikzpicture}[overlay,remember picture]
        \end{tikzpicture}
        
    \caption{{Time averaged ${\bf E} \times {\bf B}$ shearing rate $\langle \omega_{E \times B} \rangle_t$ (left) and  $\langle v_{E \times B}\rangle_t$ due to the zonal electrostatic potential component $\bar{\phi}$   (right). Comparison between  hybrid and fully kinetic models. Gradient-driven ES simulations.}}
    \label{ExB_shear_fullykin_vs_hyb}
\end{figure}

The sharp variations in $\chi$ {(here  computed as an effective $\chi$, see equation \ref{chi_computation})} are mostly attributable to strong corrugations of the temperature gradient. Indeed, the logarithmic temperature gradients  after the turbulence saturation { have evolved substantially away from their initial values}, as seen in  Figure \ref{temperature_gradients}. As we can see, the significant drop in the transport coefficients is indeed due to the increase of temperature gradients that can suggest the onset of a transport barrier since { in that region transport stiffness is  reduced}.
Interestingly, the fully-kinetic model seems also to allow for an internal $s \simeq 0.4$ transport barrier for both ions and electrons, while the hybrid model misses this effect for the ions.  One can notice that the peaks for the $R/L_T$ are slightly shifted in the hybrid model compared to the fully-kinetic model. The $R/L_T$ corrugation is associated with a strong negative peak of the time averaged $\omega_{E\times B}$ at $s\sim 0.42$, which is present in the fully-kinetic model but not in the hybrid model. See Figure \ref{ExB_shear_fullykin_vs_hyb}.

\re{\cancellable{This can be the result of a shift that is observed in electron $v_\parallel$, shown in Figure \ref{v_parallel}. We believe that in both cases this $v_\parallel$ structure, more precisely its radial shear, plays a major role in dropping the transport coefficients and in setting the high RLT. }}

\subsection{Electromagnetic} \label{EM_GD}
\begin{figure}[htbp]
    \centering
        \includegraphics[width=0.8\textwidth]{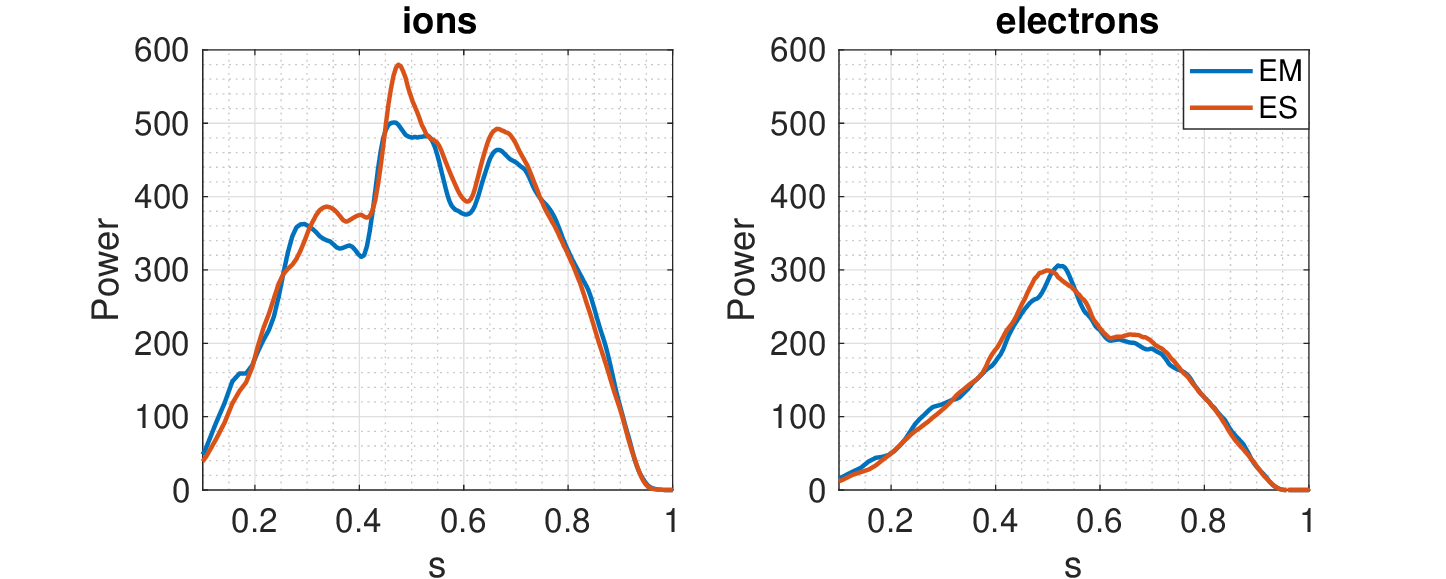}
        \begin{tikzpicture}[overlay,remember picture]
        \end{tikzpicture}
        
    \caption{{Heat power comparison between  electrostatic (red) and electromagnetic with $\beta = 10^{-4}$ (blue) simulations. Gradient-driven  simulations. }}
    \label{power_ES_EM}
\end{figure}

For the electromagnetic (EM) simulation we employ a small $\beta$, $\beta = 4\cdot10^{-4}$ { (defined with the pressure at  the $q_{\text{min}}$ position, i.e. $s=0.52$, and $B$ on axis)},  to avoid affecting linear properties of the ITG (such as the decrease of the growth rate with  $\beta$ before hitting the ITG-KBM transition). 
The dynamics is very similar to the purely electrostatic  case, and the fluxes  comparison is shown in Figure \ref{power_ES_EM}. Even though the results are qualitatively identical, with the corrugations happening at the same radial location, a tiny quantitative difference is present and it can suggest that for the ion channel  EM effects can have a small stabilizing effect even at very low $\beta$. However, the difference is so small that it would be undetectable in an experiment, and it is difficult to argue that such a difference is physical rather than numerical. 

\begin{figure}[htbp]
    \centering
        \includegraphics[width=0.8\textwidth]{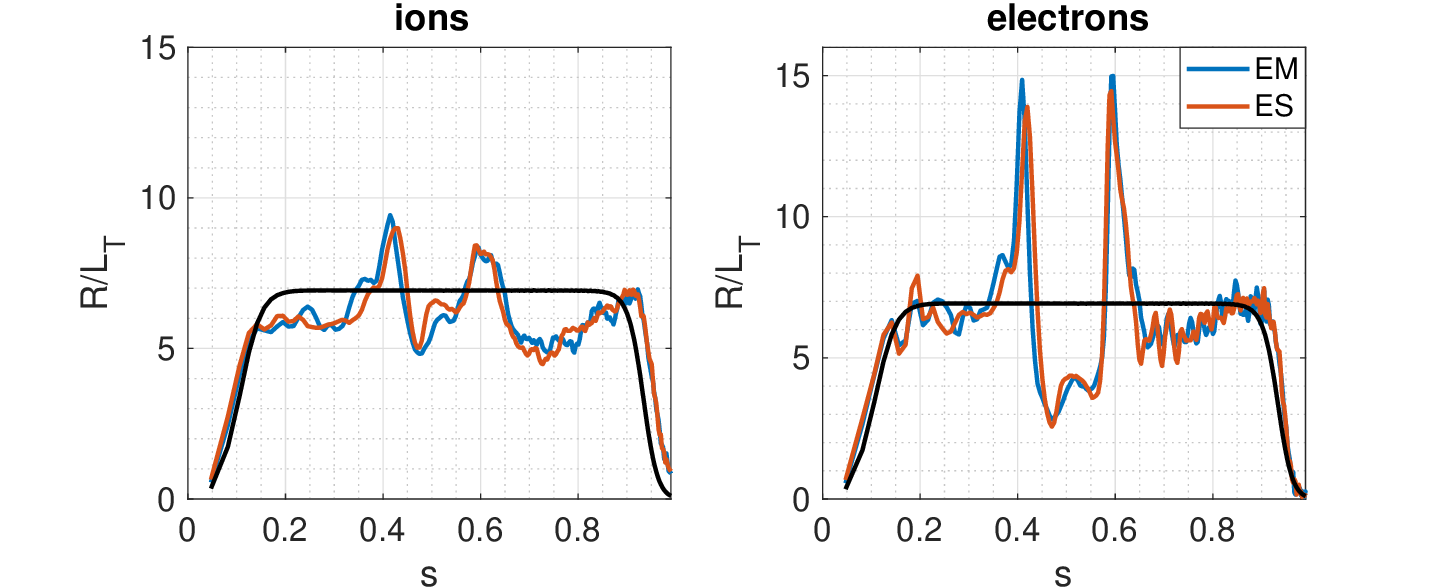}
        \begin{tikzpicture}[overlay,remember picture]
        \end{tikzpicture}
        
    \caption{Comparison of evolved  $R/L_T$ between electrostatic (red) and electromagnetic  with $\beta = 10^{-4}$ (blue) simulations.  The black line represents the initial gradients. Gradient-driven  simulations.}
    \label{RLT_ES_EM}
\end{figure}

Since temperature-gradient-driven simulations in a global GK code still allow for some temperature variations, it is important to  also compare  the relative final gradients. The comparison is shown in Figure \ref{RLT_ES_EM}. Once again, the two simulations look very similar, {and one can observe that the steepening of the gradients is much stronger for electrons than for ions (similarly to what happens with the hybrid model)}. 
The corrugation in the electron channel is impressive for both, and there is only a small relative difference between the two cases.

\begin{figure}[htbp]
    \centering
        \includegraphics[width=0.8\textwidth]{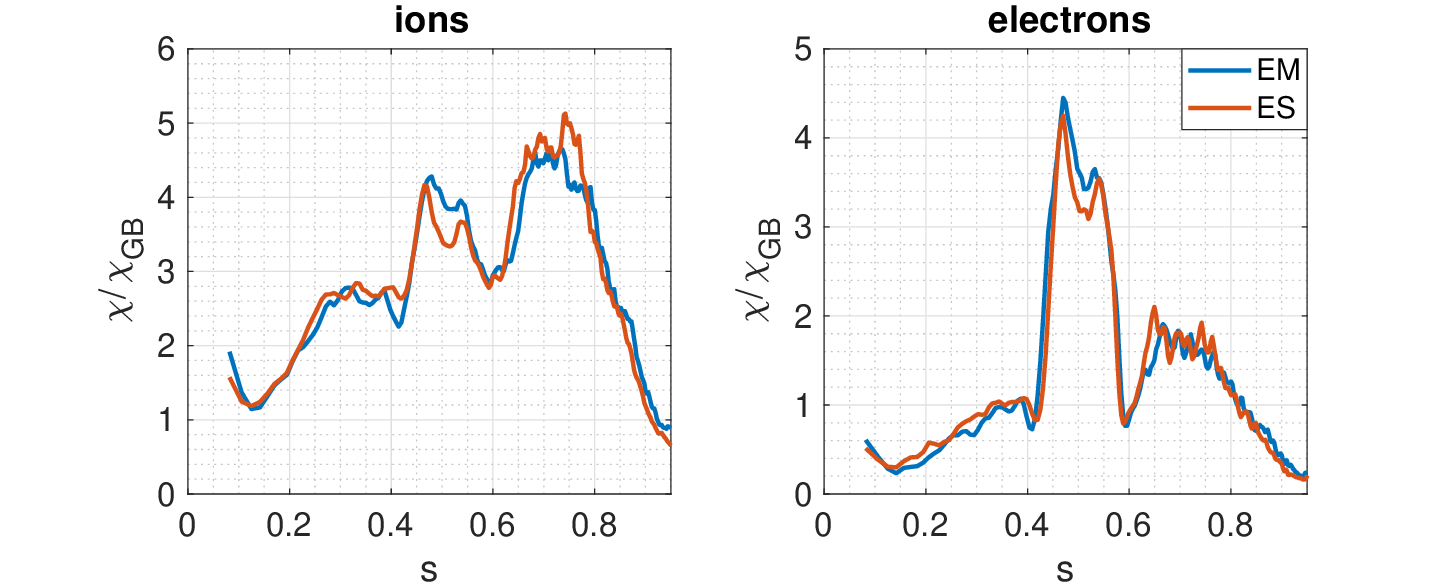}
        \begin{tikzpicture}[overlay,remember picture]
        \end{tikzpicture}
        
    \caption{{Comparison of the effective heat conductivity $\chi_{\mathrm{eff}}$  between  electrostatic (red) and electromagnetic with $\beta = 10^{-4}$ (blue) simulations. Gradient-driven simulations.}}
    \label{chi_EM_ES}
\end{figure}

The ion temperature gradient corrugation is {less pronounced than for electrons}, but in the core is slightly stronger for the EM case ($s \simeq 0.4$).
{This strongly corrugated temperature profile  {results from strongly corrugated transport}, shown in {the form of an effective $\chi$ in} Figure \ref{chi_EM_ES}, since the heat flux {does not show radial corrugations}. The continuity of heat flux for strongly discontinuous gradients is actually a sign of transport non-locality: {we note that $\chi_{\mathrm{eff}}$ has {\em maxima} where $R/L_T$ has {\em minima} (and vice-versa). } }

\begin{figure}[htbp]
    \centering
        \includegraphics[width=0.5\textwidth]{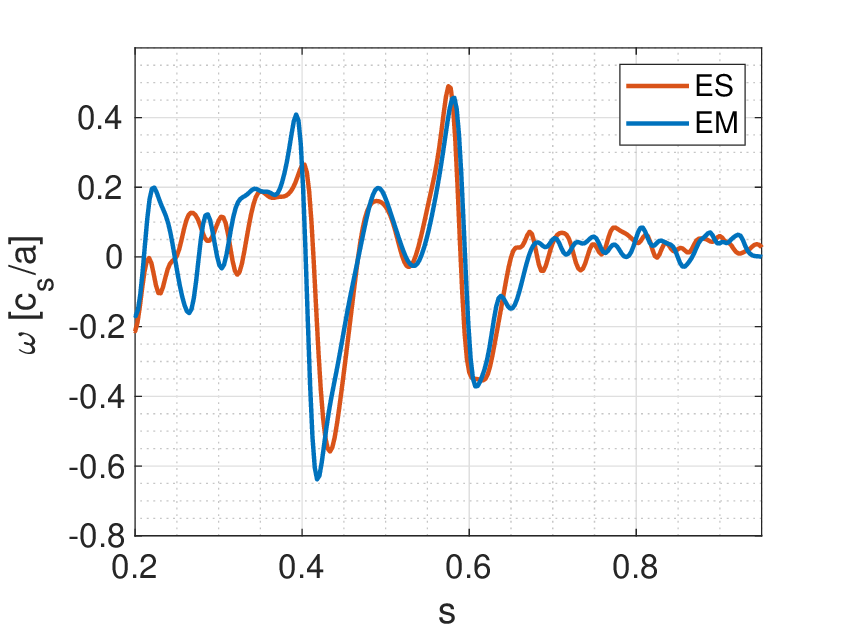}
        \begin{tikzpicture}[overlay,remember picture]
        \end{tikzpicture}
        
    \caption{{Comparison of the time averaged ${\bf E} \times {\bf B}$ shearing rate $\langle \omega_{E \times B} \rangle_t$  between  electrostatic (red) and electromagnetic with $\beta = 10^{-4}$ (blue) simulations. Gradient-driven  simulations.}}
    \label{shearing_rate_EM_ES}
\end{figure}

The $\omega_{E\times B}$, which is shown in Figure \ref{shearing_rate_EM_ES}, is also similar between the EM and the  ES cases. The most relevant (still small) difference is around $s \simeq 0.4$ which may explain the lower power for the EM case.

As usual the corrugation in the shearing rate $\omega_{E\times B}$ matches the corrugation observed in the $R/L_T$ profiles (Figures \ref{RLT_ES_EM}-\ref{shearing_rate_EM_ES}).

\begin{figure}[htbp]
    \centering
        \includegraphics[width=0.7\textwidth]{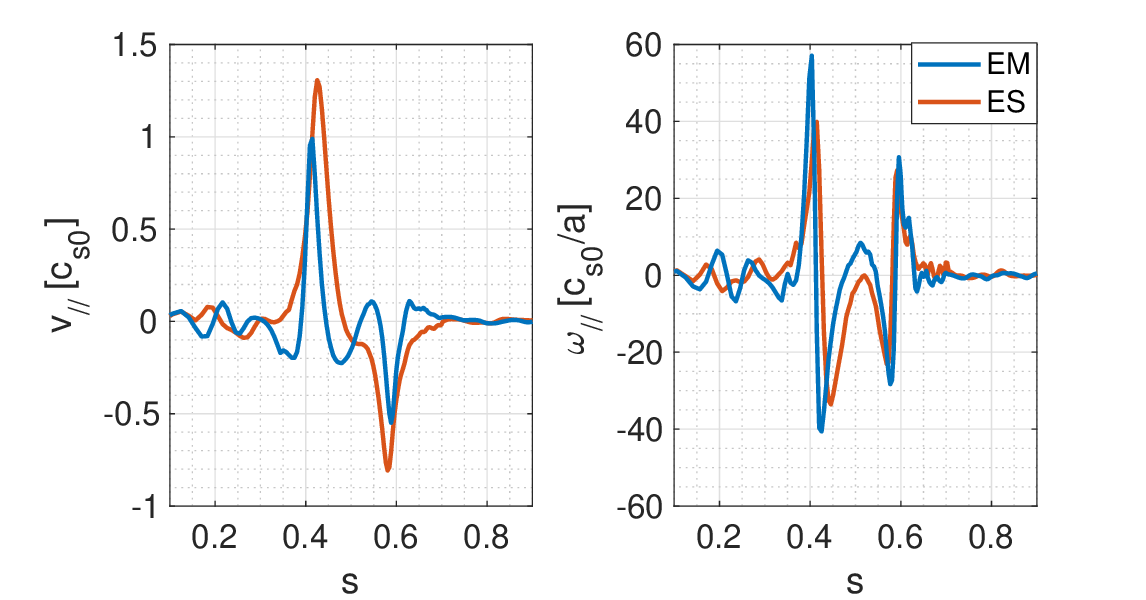}
        \begin{tikzpicture}[overlay,remember picture]
        \end{tikzpicture}
        
    \caption{{Comparison of parallel velocity  between  electrostatic (red) and electromagnetic with $\beta = 4\;10^{-4}$ (blue) simulations. Gradient-driven  simulations.}}
    \label{vparal_ES_EM}
\end{figure}

As we already stressed, the strong $\omega_{E\times B}$ corrugation comes {together} with a strong layer of zonal electron parallel velocity, {and hence a current}. The structure of zonal current and its shear  are displayed in Figure \ref{vparal_ES_EM}.
It is interesting to note that $v_\parallel$ saturates at lower values for the EM case. 
This  may be due to the fact that, in a purely electrostatic simulation, the current does not provide a non-linear feedback to the system, while in the EM simulation, it does through the perturbed vector potential.

Furthermore, while in the ES simulation there is a smooth transition between positive and negative peaks of $v_\parallel$, in the EM simulation there is a part of the domain where the $v_\parallel$ goes up again before the negative peak.  This structure in velocity results in a region where the $v_\parallel$ shear $\omega_\parallel$ goes up to positive values, whereas in the ES case,  the maximum  $\omega_\parallel$ in region $s \in [0.4,\;0.6]$ is 0. This difference is displayed in Figure \ref{vparal_ES_EM}-right panel.\\

\begin{figure}[htbp]
    \centering
        \includegraphics[width=0.5\textwidth]{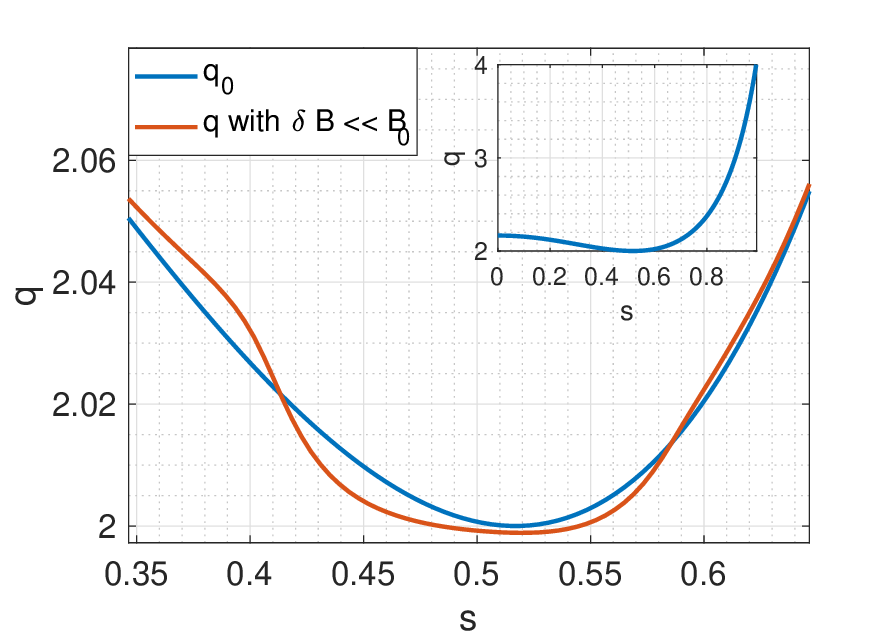}
        \begin{tikzpicture}[overlay,remember picture]
        \end{tikzpicture}
        
    \caption{{Safety factor modification due to the zonal component of $A_\parallel$ (EM simulation). In red, the q-profile including turbulence induced modifications; in blue, the initial one. In the subplot the complete initial q-profile is shown. Gradient-driven  simulations.}}
    \label{modified_q}
\end{figure}

The inclusion of the Amp\`ere equation  also leads to another effect: an effective modification of the $q$-profile. This effect, extensively discussed in Ref. \cite{Volcokas_Arxive2024_magn}, is shown in Figure \ref{modified_q}. A persistent dipole structure in the axisymmetric component of $A_\parallel$ develops self-consistently during turbulence saturation and stays on for  the whole simulation duration, leading to {a modification of the safety factor profile}. This long-living zonal $A_\parallel$ is a {result of}  the $v_\parallel$ structure generated by the turbulence shown in Figure \ref{vparal_ES_EM}.  
The emerging $A_\parallel$ tends to flatten out the q-profile in the region around the minimum and locally increases  the magnetic shear near the peaks of $v_\parallel$. This might help to further stabilize the ITGs and it could explain the small difference between the ES and EM in Figure \ref{power_ES_EM}-a. {While in the flux-tube setup  the stabilizing effect is clear \cite{Volcokas_Arxive2024_magn}, here we do not see a particularly strong stabilizing effect on the final heat flux: {the differences shown in Figure \ref{power_ES_EM} are most probably undetectable experimentally.}}  Actually, a difference {shows up} at the beginning when gradients are still evolving, but as they  reach the final state, their contribution to turbulence stabilization seems to be more important and the two cases converge to almost the same quasi-steady state.

However, the scans performed in \cite{Volcokas_Arxive2024_magn} were conducted using different parameters; {thus a direct one-to-one comparison should be avoided}.

The modified q-profile shown in Figure \ref{modified_q} has been computed using  equation  \ref{safety_2}.

\section{Flux-driven simulations} \label{fluxdriven_section_results}

{The flux-driven approach  allows one to stress how the temperature profiles start to corrugate and to see the emergence of a transport barrier}. 
Moreover,  it enables observation of  how the system reacts when there are no constraints on the temperature profiles ({as opposed to constraints that are imposed in gradient-driven simulations to keep the temperature gradients close to the initial ones}). \re{\cancellable{As we will see, the constraints in temperature may lead to incapability of the system to evolve as it would do in a flux driven simulation.}}

\begin{figure}[htbp]
    \centering
        \includegraphics[width=0.5\textwidth]{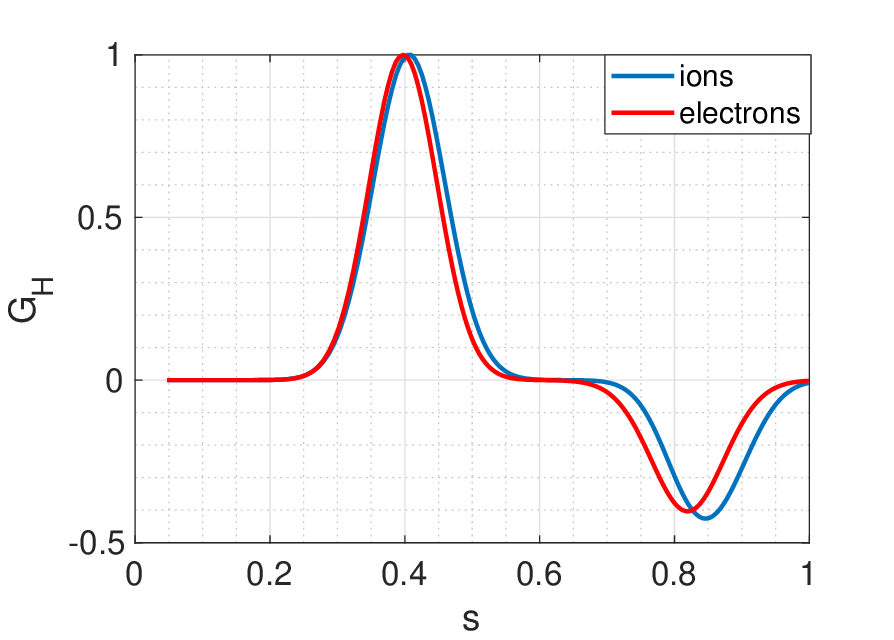}
        \begin{tikzpicture}[overlay,remember picture]
        \end{tikzpicture}
        
    \caption{{Input power shape for the flux-driven simulations.}}
    \label{flux_shape}
\end{figure}

To perform flux-driven simulations, a certain input power must be provided. The flux-surface-average heating operator is:

\begin{equation}
    S(s,E) = \gamma_h \,G_H(s) \, \frac{1}{T(s)/m} \left ( \frac{E}{T(s)/m} - \frac{3}{2}   \right ) f_L (n(s),T(s),E) + S_H^c,
\end{equation}

where $n(s), \; T(s)$ are the flux-surface-averaged density and temperature profiles at the {beginning of the simulation}, $G_H$ is the radial profile for the heat source used for localized heating, and the constant $\gamma_h$ is the source strength. The first term is proportional to the temperature derivative of a reference local Maxwellian function $f_L, \, \frac{\partial f_L}{\partial T}$. The consequence is that there is no density source in the infinite marker limit, since $\int dV \partial_T f_L = 0$. The last term, $S_H^c$,  is a correction term that aims at ensuring flux surface averaged  conservation of parallel momentum, zonal flows and density (since we are not in the infinite markers limit) up to machine precision. We stress that there is no restriction in having $S$ everywhere positive, it can be used also to cool-down the system (i.e. to act as a sink). \cancellable{This is indeed what we do.} Since flux-driven simulations without any {a priori knowledge} of final profiles corresponding to a certain input power are prohibitive at this stage, we exploit the information coming from gradient-driven simulations. 
From gradient-driven simulations we determine the heating power that the system requires in order to maintain the temperature relatively close to the initial one. Thus, we take this profile and  smooth it in order to have a more realistic heat source. The heat source profiles $G_{{i,e}}$ used for our analysis are shown in Figure \ref{flux_shape}, and the coefficients $\gamma_{h_{i,e}}$ are chosen to have the same integrated power source (up to $s=0.6$) as the gradient-driven case. We chose to not perfectly balance out the positive and the negative part of the source. This might indeed cause some problems during the beginning of the simulation if the sink removes energy before turbulence starts to transport energy from the core to the edge. Thus, to totally balance out the slightly positive integrated source, we use a buffer that removes all the remaining fluctuations at the edge. This is done with the following Krook operator:
\begin{equation}
    S_B = -\gamma_B \left ( \frac{s -s_B}{1-s_B}  \right )^4 \delta f \quad , \quad \text{for } s > s_B \quad .
\end{equation}
This operator  is particularly useful since it is not fixed but its strength changes according to the magnitude of the fluctuations ($\delta f$). In our simulations $s_B = 0.95$. 

To show what we believe is the onset of the ITB and how this is related once again to $q_{\text{min}}=2$, we compare two different safety factor profiles as in section \ref{hybrid_section_results} (i.e., one with $q_{\text{min}}=2$ and the other with $q_{\text{min}}=2.03$). We stress that with limited computational resources at hand it is particularly challenging  to achieve a quasi-steady state condition where the source completely balances out the losses due to transport. However, the results are robust and the qualitative picture is clear. \\

\subsection{Importance of self-interaction for ITB formation: $q_{min}= 2$ versus $q_{min}= 2.03$}
\begin{figure}[htbp]
    \centering
        \includegraphics[width=1\textwidth]{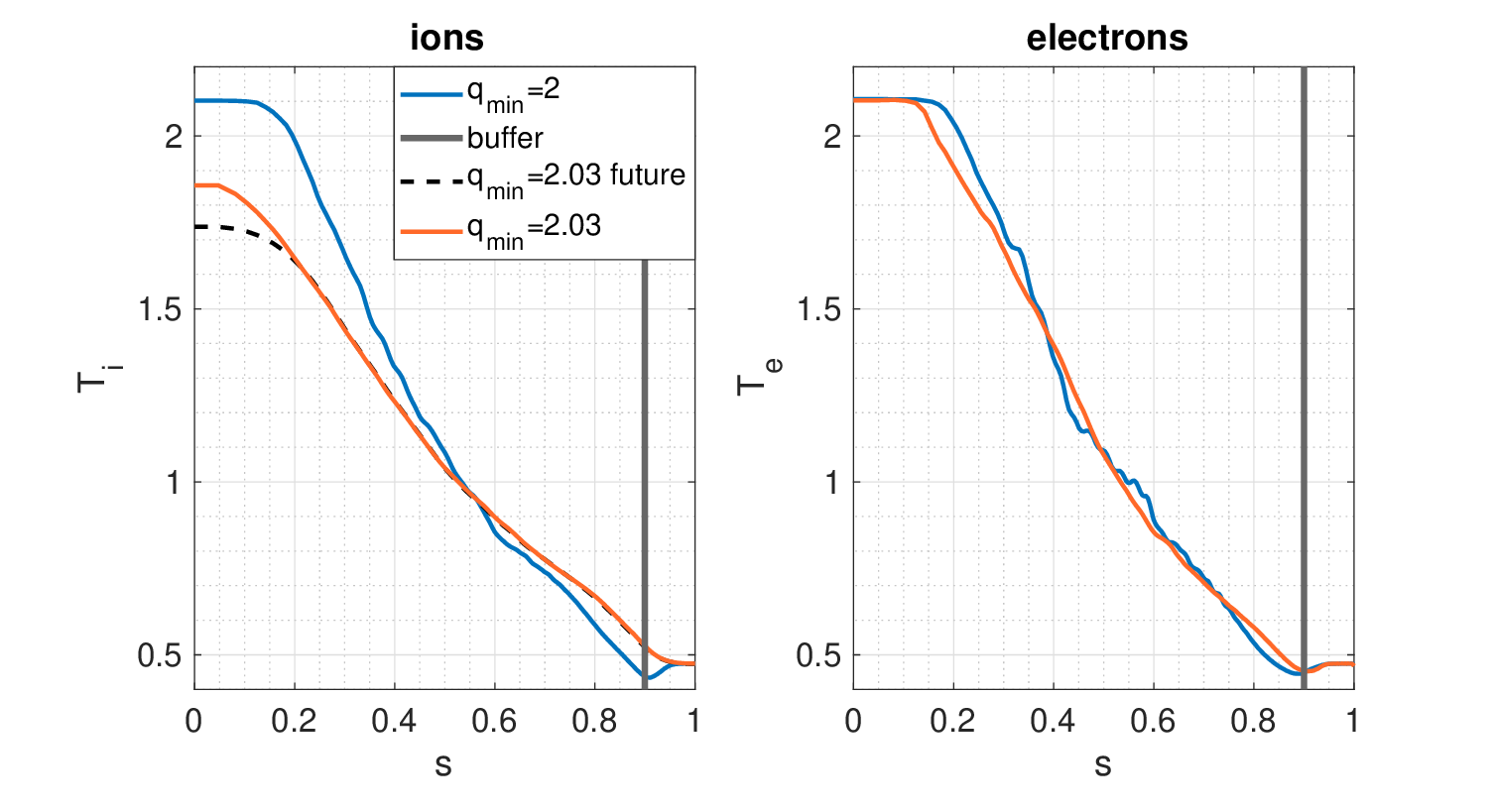}
        \begin{tikzpicture}[overlay,remember picture]
        \end{tikzpicture}
        
    \caption{{{Final} temperature profiles for ions (left) and electrons (right) for the cases with $q_{\text{min}}=2$ (blue) and $q_{\text{min}}=2.03$ (red and dashed black). The red curve represents the stage where we stopped the simulation, the dashed curve is the core prediction after the smoothing of $R/L_T$ (see  Figure  \ref{RLT_q2_qbar2}). Flux-driven EM simulations. }}
    \label{comparison_qmin2_qbar2}
\end{figure}

\begin{figure}[htbp]
    \centering
        \includegraphics[width=1\textwidth]{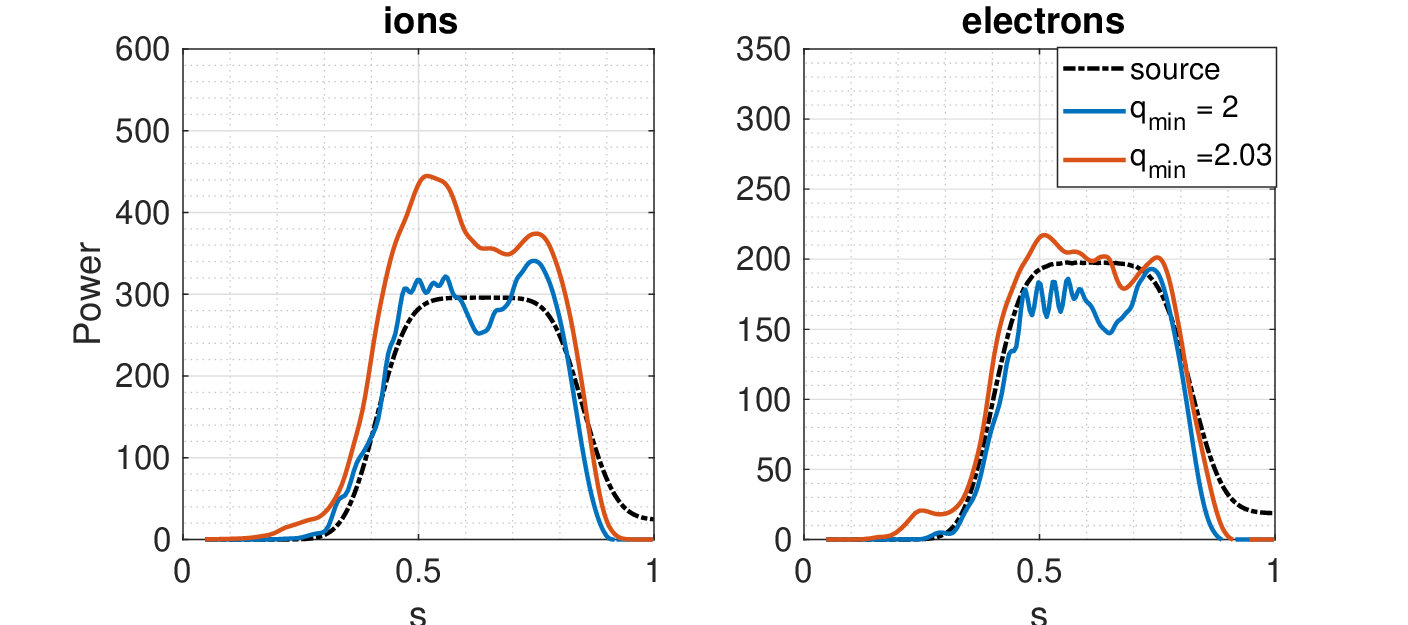}
        \begin{tikzpicture}[overlay,remember picture]
             \node at (-1., 6.8) {\textbf{a)}}; 
             \node at (5.5, 6.8) {\textbf{b)}}; 
        \end{tikzpicture}
        \caption{{Power balance: radial integral of the input power (dashed line) versus the final (at the stage we stopped the simulation) losses. Ions left, electrons right.  Blue $q_{\text{min}}=2$, red $q_{\text{min}}=2.03$. Flux-driven EM simulations.}}
    \label{sources_vs_losses}
\end{figure}

In Figure \ref{comparison_qmin2_qbar2} the ion and electron temperatures are shown for the two simulations. The first thing to notice is that the difference between the two configurations is evident for the ions, while electrons feature similar profiles. This is actually due to the fact that the prominent instability is the ITG (we remind that the density gradients are zero). We display an average over the last $30\, a/c_s$ of simulations. The $q_{\text{min}}=2.03$ case has not reached {quasi steady-state} yet {(although from a power balance analysis, shown in Figure \ref{sources_vs_losses}, it is approaching it)}. In the plot relative to the ions, there is an additional black-dashed line for the $q_{\text{min}}=2.03$ case. This line is an extrapolation of the final core profile {based on the fact that the system has not reached {quasi steady-state} in the core. Indeed, a power balance analysis shows that (Figure \ref{sources_vs_losses}, left) the turbulent power losses there (red curve) exceed the input power (black dashed curve). The ion temperature is therefore expected to further decrease in the core.} 
We stress that this prediction is conservative (i.e. it does not consider that  the  $R/L_T$ at $s>0.23$ will also go down): we expect the real difference to be larger due to a power balance analysis. 
Indeed the power balance (heat power given versus heat power lost through turbulent transport) shown in Figure \ref{sources_vs_losses} can be used to assess how far we  are from quasi steady-state and in which direction the profiles are expected to evolve.

\begin{figure}[htbp]
    \centering
        \includegraphics[width=0.5\textwidth]{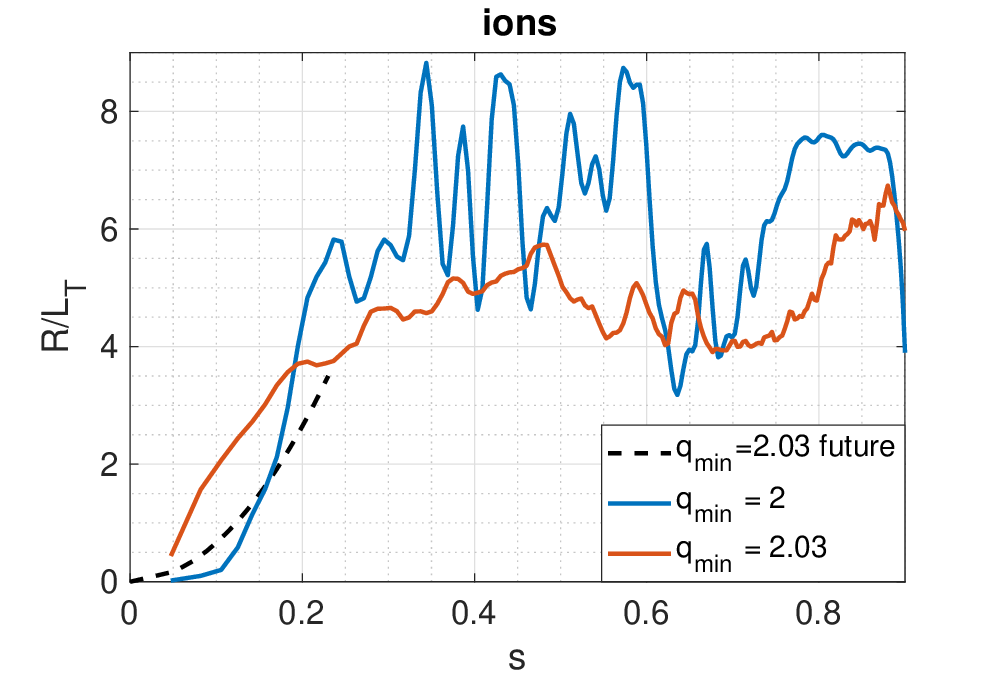}
        \begin{tikzpicture}[overlay,remember picture]
        \end{tikzpicture}
        
    \caption{Logarithmic gradient profiles for the $q_{\text{min}}=2$ (blue) and $q_{\text{min}}=2.03$ (red and dashed black). {The black curve is an extrapolation to the future as the temperature for $s \in [0, 0.23]$ is still dropping (since losses are larger than input power) and typical parabolic profiles in the very core are expected (as  for the $q_{\text{min}}=2$ case). Flux-driven EM simulations.} }
    \label{RLT_q2_qbar2}
\end{figure}

In the plot, the black dashed curve represents the radially integrated source of Figure \ref{flux_shape} while the continuous curves are the power losses through turbulent transport. At the steady state, by definition, the continuous curves and the dashed ones must coincide. However, at this stage of the simulations, especially for the $q_{\text{min}}=2.03$ case the losses are more than the total injected power. In particular, referring to the extrapolation of the ion logarithmic gradient (black curve of Figure \ref{RLT_q2_qbar2}) and consequent reconstruction of ion temperature (black curve of Figure \ref{comparison_qmin2_qbar2}), it is evident how we can justify our assumption: while  for the case $q_{\text{min}}=2$ the power balance is satisfied, for the case $q_{\text{min}}=2.03$ in range $s \in [0.2, 0.3]$ the power losses are $\simeq 50$ ({ORB5 units}) for the ions and $\simeq 20$ for the electrons (while the input powers are $\simeq0.01$ and $\simeq0.005$ respectively). 
{Despite the ion temperature already showing a significant difference in the two cases, the losses due to transport are still larger than the sources for the $q_{\text{min}}=2.03$ case.}

 On longer time scales, this will lead to  a  {\itshape{global}}    (and not just in the region $s < 0.23$ as in our black extrapolation) drop  in the logarithmic gradient for the  $q_{\text{min}}=2.03$ case, making our extrapolation in Figure \ref{comparison_qmin2_qbar2} an upper {(optimistic)} estimation for the $q_{\text{min}}=2.03$ ion temperature profile {(meaning that the converged ion temperature profile will drop even below the dashed line)}. 
 Finally, we want to stress that, since the initial profile was the same for the two configurations, having a pre-defined positive and negative source (sink) leads to higher  temperature { at the entrance of the buffer region near the edge ($s\approx 0.9$)} for the case with higher transport coefficients. Renormalizing the final temperature at the buffer position, the core ion temperature differences would be much larger.

\begin{figure}[htbp]
    \centering
        \includegraphics[width=1\textwidth]{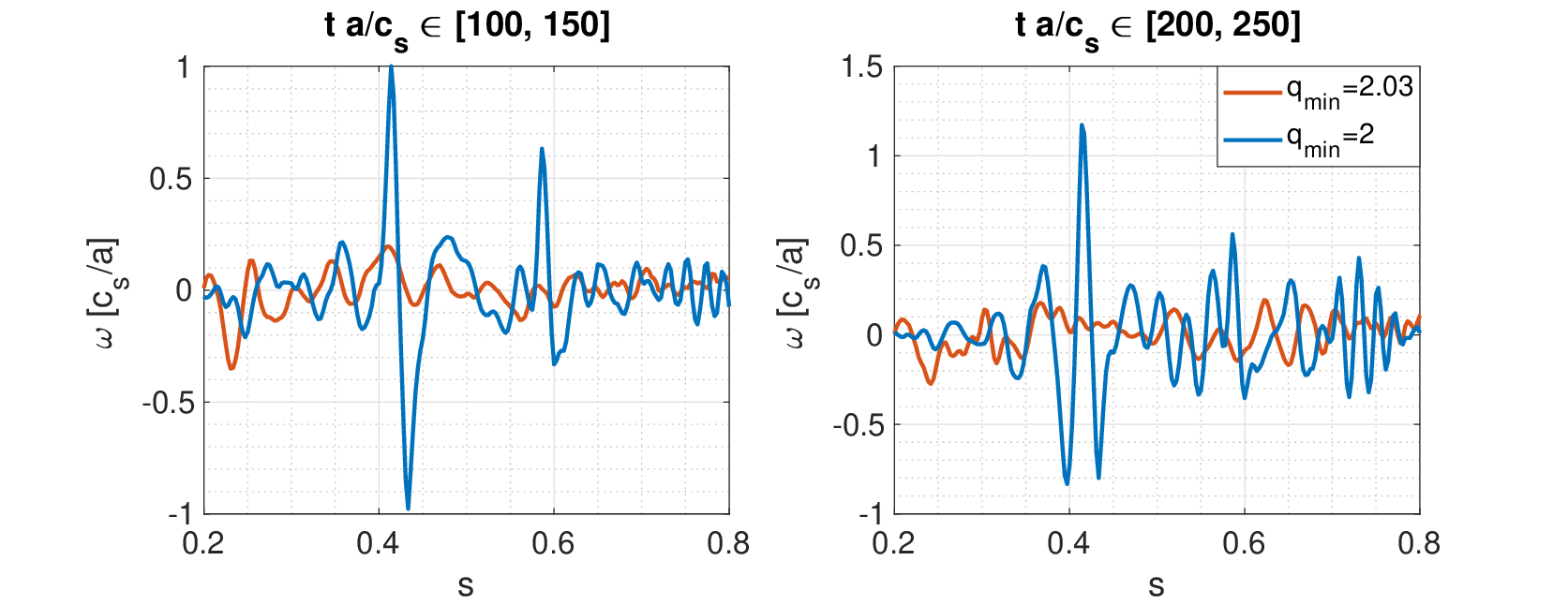}
        \caption{{ Time averaged ${\bf E} \times {\bf B}$ shearing rate $\langle \omega_{E \times B} \rangle_t$ , averaged in $t \, c_s/a \in [100, 150]$ (left) and $t \, c_s/a \in [200, 250]$ (right) for the $q_{\text{min}}=2$ and  $q_{\text{min}}= 2.03$ cases (blue and red, respectively). Flux-driven EM simulations.}}
    \label{exb_shear}
\end{figure}

 The strong stabilization in the ion channel of the $q_{\text{min}}=2$ case can be, at least partially, explained by the strong  $\omega_{E\times B}$, shown in Figure \ref{exb_shear}. The figures show the $\omega_{E\times B}$ averaged between two different time intervals for the cases $q_{\text{min}}=2$ and  $q_{\text{min}}=2.03$. It is evident that  {after the nonlinear saturation} two strong $\omega_{E\times B}$ layers develop for $q_{\text{min}}=2$ that are able to suppress turbulence correlation. Then a much { more strongly corrugated }  $\omega_{E\times B}$ develops everywhere. However, what happens during the first phase $t \, c_s/a \in [100, 150]$ is crucial to suppress turbulence.
 
 Indeed, in this case the $\omega_{E\times B}$  is strongly radially corrugated {ranging within } only a few $\rho_i$ from $ 0.5 \, c_s/a$ to $ -0.3 \, c_s/a$ in the region $s \simeq 0.6$ and from  $ 1 \, c_s/a$ to $ -0.8 \, c_s/a$ in the region $s \simeq 0.4$, i.e. in the two radial locations where the transport barriers develop. This value is much larger than the global linear growth rate $\gamma$ of the instability (extracted in the linear phase of the simulation) that is $\gamma \simeq 0.32 \,c_s/a$: this leads to a significant stabilization effect since the eddies are sheared and broken faster than they are created.
Furthermore, this effect is also increased by the fact that around the transport barriers the $\omega_{E\times B}$ changes its sign remaining large in amplitude: when $\omega_{E \times B}$ swaps sign, it means that the zonal $v_{{E}\times B}$ has the structure of a parabolic flow. In the field of dynamical system theory it has been shown that parabolic flows can be associated, from a mathematical point of view, to parabolic transport barriers \cite{Farazmand_PhyD2014} (here to be intended as transport barriers for material points that follow a given trajectory resulting from the integration of a given velocity field). It is not straightforward to extrapolate the parabolic barrier to our system where the velocity field results from turbulence, but the fact that the zonal component of $\phi$ in quasi-steady state does lead to quasi steady contribution to the velocity field that can finally be linked to the parabolic barriers described in  \cite{Farazmand_PhyD2014}. These kinds of transport barriers have also been proven  to persist  under certain perturbations, in particular when a small noise is added to a deterministic flow \cite{Balasuriya_pre2018}. In our case, we think that this can partially explain the barrier provided that the steady zonal $v_{{E}\times B}$ is substantially larger than the fluctuating $v_{\Tilde{E}xB}$, that is clearly valid in the $q_{\text{min}}=2$ case around the transport barriers.

\begin{figure}[htbp]
    \centering
        \includegraphics[width=0.9\textwidth]{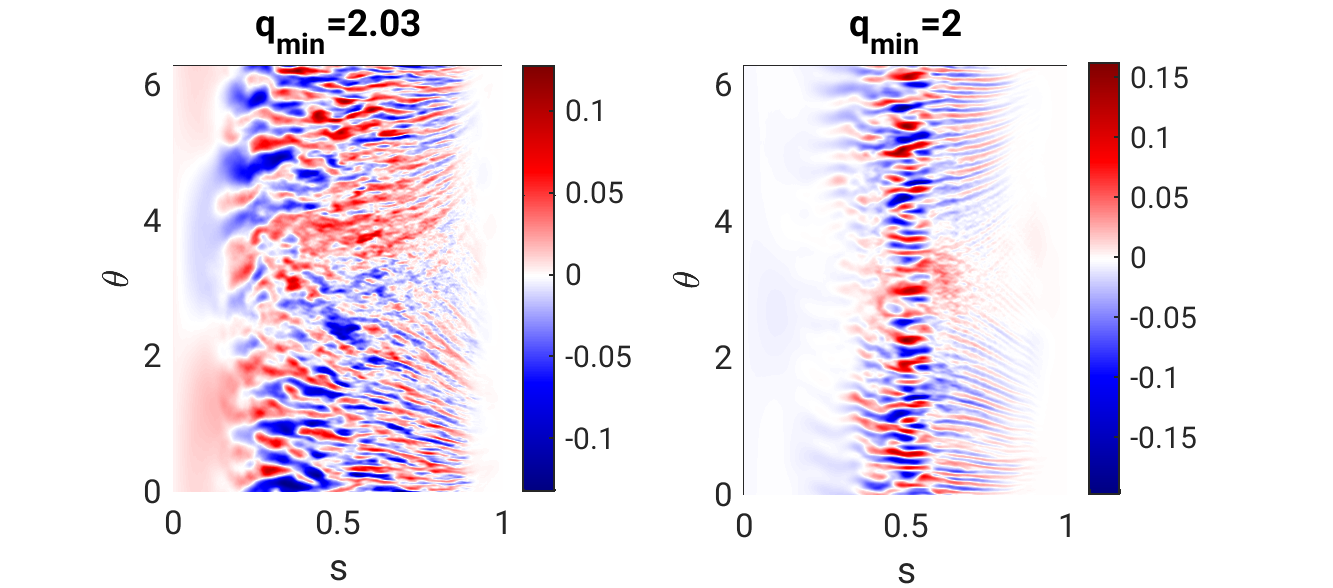}
        \caption{{Time snapshot of the non-zonal electrostatic potential $\tilde{\phi} = \phi - \bar{\phi}$ (with $\bar{\phi}$ the zonal component) at the toroidal angle $\zeta=0$ and at $t \,c_s/a = 150$ for the $q_{\text{min}}=2.03$ (left) and $q_{\text{min}}=2$ (right) as a function of the radial and poloidal coordinates. Flux-driven EM simulations.}}
    \label{phi_pert}
\end{figure}

\begin{figure}[htbp]
    \centering
        \includegraphics[width=0.5\textwidth]{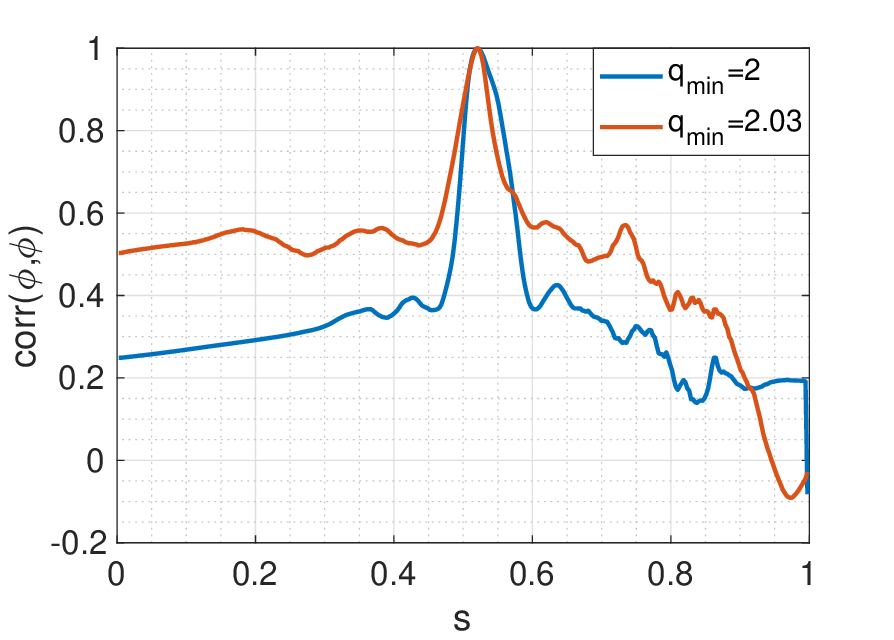}
        \caption{{ Non-zonal electrostatic potential $\tilde{\phi} = \phi - \bar{\phi}$ (with $\bar{\phi}$ the zonal component) correlation with respect to the point $s=0.52$ for the $q_{\text{min}}=2, \; 2.03$ cases (blue and red, respectively). Flux-driven EM simulations}}
    \label{correlation}
\end{figure}

{Thus, this effect spatially de-correlates the fluctuations in the regions separated by the peaks and troughs of  $\omega_{E\times B}$, leading to smaller eddies and reduced transport.}
This can be confirmed looking at a temporal snapshot of $\Tilde{\phi} = \phi(s,\theta,0) - \bar{\phi}$ {(with $\bar{\phi}$ zonal component of $\psi$)} , shown for the two cases in Figure \ref{phi_pert}. It is evident that in the $q_{\text{min}}=2.03$ case the eddies are radially elongated and the regions $s \in [0, 0.4], \; s \in [0.4, 0.6], s \in [0.6, 1]$ are connected, while in the  $q_{\text{min}}=2$ case the three regions are not connected. To better show this feature, the time-correlation 
\begin{equation}
    \text{corr}(\Tilde{\phi}_{s1},\Tilde{\phi}_{s2}) = \langle \Tilde{\phi}_{s1}\,\Tilde{\phi}_{s2}\rangle_t / (\sigma_{s1} \, \sigma_{s2})
\end{equation} \label{correl}

between $\Tilde{\phi} $ at $s=0.52$ and  the other radial points is shown in Figure \ref{correlation}. The correlation of fluctuations between this region and the regions $s>0.6$ and $s<0.4$ is strongly reduced for the $q_{\text{min}}=2$ case compared to the $q_{\text{min}}=2.03$ case.

{To further confirm the prominent role of zonal flows, after reaching quasi steady-state we removed from the $q_{\text{min}}=2$ simulation the toroidal mode $n=0$ and two things happened: a) the heat flux increased by a factor of 3; b) the 2D plot of $\tilde{\phi}$  became  similar to the one of the $q_{\text{min}}=2.03$ case.}

\begin{figure}[htbp]
    \centering
        \includegraphics[width=0.6\textwidth]{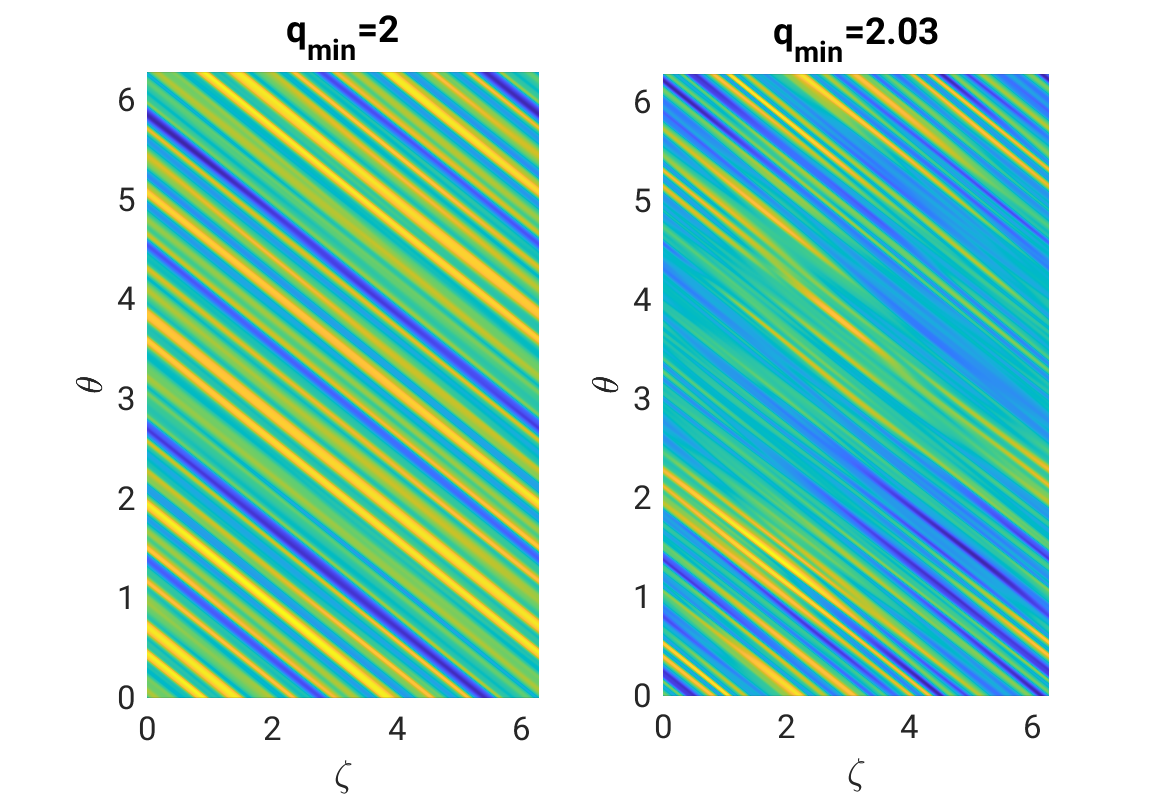}
        \caption{{Contours of $\tilde{\phi} = \phi - \bar{\phi}$ (with $\bar{\phi}$ the zonal component) at $s=0.52$ (corresponding to the $q_{\text{min}}$ location)  as a function of $\theta$  and $\zeta$, poloidal and toroidal angles respectively. Left panel: $q_{\text{min}}=2$ case; right panel: $q_{\text{min}}=2.03$ case. Flux-driven EM simulations.}}
    \label{parallel_elongation}
\end{figure}

It is interesting to note that $\tilde{\phi}$ shown in Figure \ref{phi_pert}  does not seem to vary poloidally for the $q_{\text{min}}=2$ case.  This is due to turbulent eddies being able to extend much further along the magnetic field lines in low magnetic shear regions and remain highly correlated, in agreement with flux-tube results in Ref. \cite{Volcokas_Arxive2024}. In Figure \ref{parallel_elongation}  we compare $\tilde{\phi}(s_{\text{min}},\theta,\varphi)$  for  $q_{\text{min}}=2$ and $q_{\text{min}}=2.03$, and we can see that the turbulent eddies for $q_{\text{min}}=2$ case are ultra long and can bite their own tail leading to strong parallel self-interaction. The $q_{\text{min}}=2.03$ case does not exhibit the same feature.

\subsection{system size comparison}
We now analyze how the properties of the flux driven case with $q_{\text{min}}=2$ change with the ratio $\rho_L/a$, i.e. with $\rho^*$. The goal is twofold:  1) to assess transport properties scaling in the presence of transport barriers; 2) to verify whether the width of the region where the safety factor flattens depends on $\rho^*$ (and thus if it is mediated by ion-scale turbulence). 

\begin{figure}[htbp]
    \centering
        \includegraphics[width=1\textwidth]{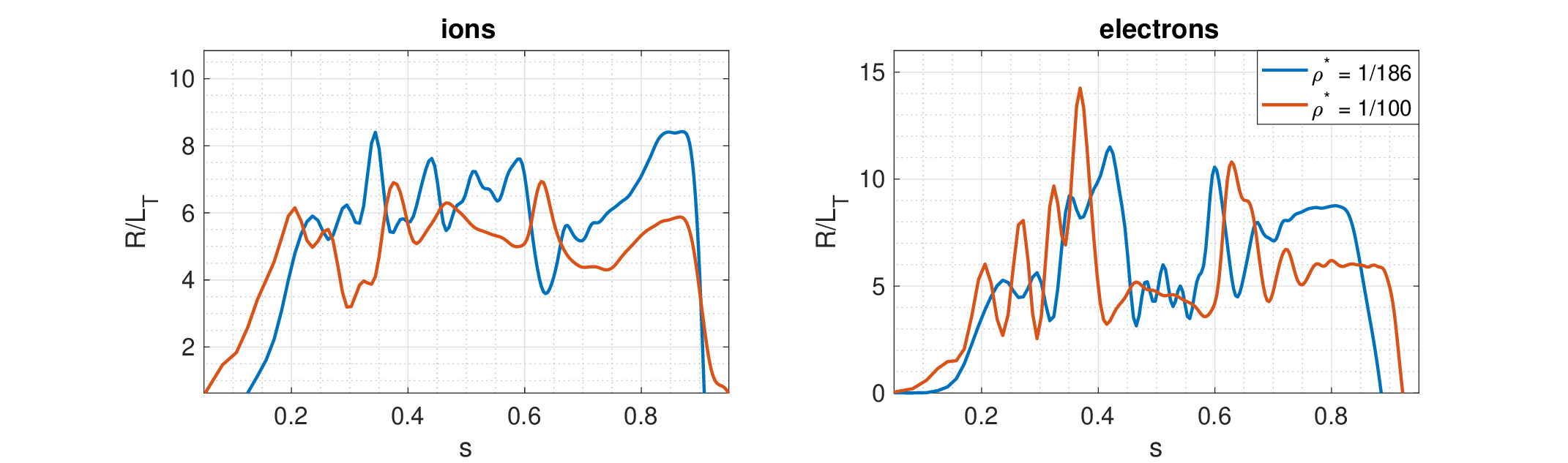}
    \caption{{$R/L_T$ comparison between the simulations with $\rho^* = 1/186$ (blue) and $\rho^* = 1/100$ (red). Flux-driven EM simulations.}}
    \label{RLT_comparison_rhostar}
\end{figure}

To this end we performed another simulation with {$\rho^* = 1/100$}, i.e. almost twice as large as the CBC $\rho^*$ employed so far and more typical of the TCV experiment. The TCV-$\rho^*$ has been heated with a power that is $\sim 80\% $ of the CBC in ORB5 units. This corresponds to also $\sim 80\% $ in SI units if one considers the plasma being in the same conditions and the only difference in $\rho^*$ is attributed to the minor radius $a$. 
The final logarithmic gradients are shown in Figure \ref{RLT_comparison_rhostar}. We now analyze the  power required to achieve such $R/L_T$ profiles and the implications.

There are different ways to consider different $\rho^*$: the simplest way would be to consider the same machine (e.g. TCV) and consider one "normal" and one cold plasma (i.e. the case with $\rho^* = 1/186$) with $T_e$ that is $(\rho^*_{CBC}/\rho^*_{TCV})^2$ colder than the usual TCV. This would give a much lower power necessary to heat the system according to the power we used in the two cases, which aligns with expectations.

\begin{figure}[htbp]
    \centering
        \includegraphics[width=1\textwidth]{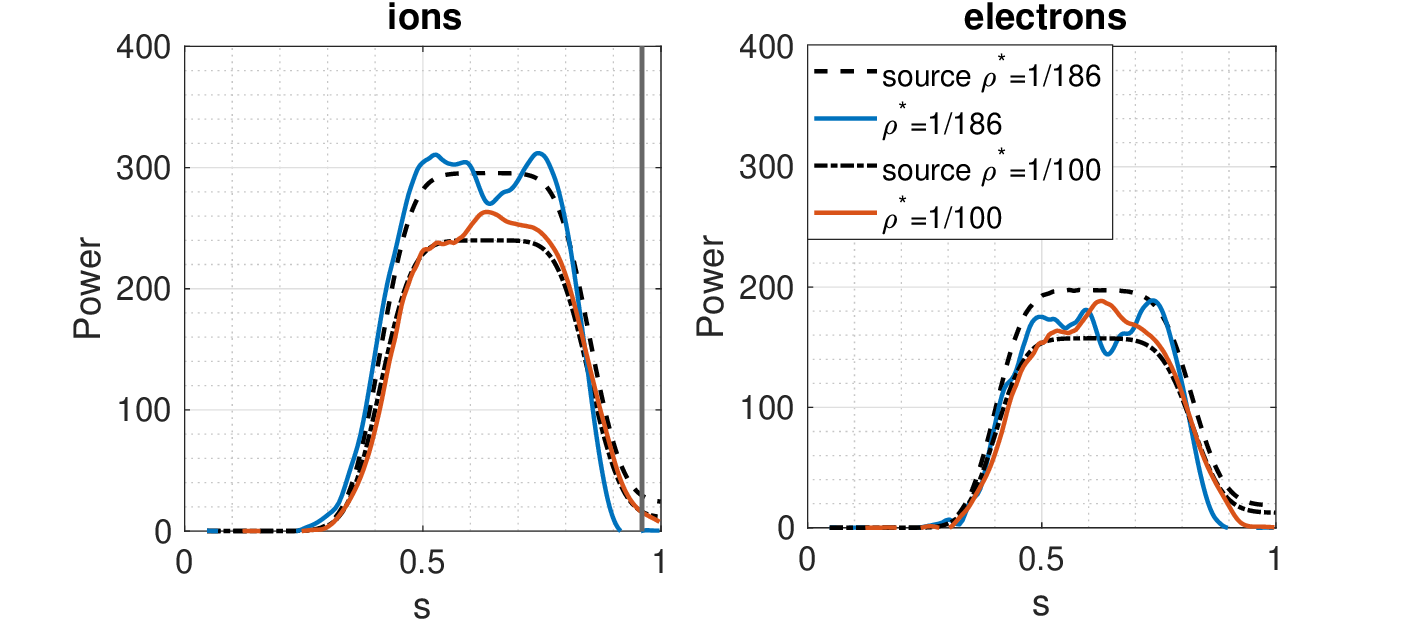}
    \caption{{Power balance analysis for  two $\rho^*$ values, TCV-like ($\rho^*=1/100$) and DIII-D-like ($\rho^*=1/186$). Flux-driven EM simulations.}}
    \label{rhostar_heat_fluxes}
\end{figure}

The opposite situation is obtained by considering the same temperature  but different machine dimensions. In this case the power ratio in SI units is 

$$
\frac{P_1}{P_2} = \frac{Q_1^{ORB5} \;n \;c_{s_1}\; T_{e_1} \;A^{ORB5}_1 \;\rho^2_{L_1}}{Q_2^{ORB5} \; n \;c_{s_2}\; T_{e_2}\; A^{ORB5}_2 \;\rho^2_{L_2}} \qquad,
$$
with $Q^{ORB5},\, A^{ORB5},\, n,\, c_s,\, T_e, $ heat flux and area of the flux surfaces (in ORB5 units), density, sound speed and electron temperature at the reference position, the subscripts (1,2) refer to the cases $\rho^* = 1/186$  and $\rho^* = 1/100$ respectively. 
Considering the same plasma temperature and density, and for simplicity the same magnetic field, the terms $c_s,\; T_e, \; \rho_L$ cancel each other and one is left with
$$
\frac{P_1}{P_2} = \frac{Q_1^{ORB5} A^{ORB5}_1 }{Q_2^{ORB5}  A^{ORB5}_2 } \qquad,
$$
with the ratio $A^{ORB5}_1 /A^{ORB5}_2 = (a_1/a_2)^2 = (\rho^*_2/\rho^*_1)^2 = 3.46$. We thus see that the ratio of powers in ORB5 units is equal to the ratio in SI units.
The power balance for the two cases is shown in Figure \ref{rhostar_heat_fluxes}. 


To summarize, according to our analysis, to heat up at the same temperature (at $s=0.52$) the DIII-D machine would require only 1.25 times the power required by TCV and it would even feature larger gradients as shown in Figure \ref{RLT_comparison_rhostar}. Targeting the same CBC gradients requires more power than the one we injected, and it would shift the ratio ${P_1}/{P_2}$ towards unity, i.e. toward gyro-Bohm scaling. Studies on transport scaling performed with adiabatic and hybrid electron models foresee an intermediate Bohm-gyro-Bohm scaling in this $\rho^*$ range \cite{McMillan_2010_PRL,Di_Giannatale_2024_ppcf}, so it seems that this particular scenario is beneficial for the scaling compared to classical monotonic safety factor profiles.

\re{\cancellable{ However, we can still compare the transport coefficients $\chi$ of the two $\rho^*$ simulations keeping in mind that heat flux and temperature gradients must increase a bit more for reaching convergence in the CBC. The $\chi$ comparison between the two $\rho^*$ simulations is shown in Figure \ref{chi_comparison_rhostar} and the final logarithmic gradients profiles are shown in Figure. }}

       

\begin{figure}[htbp]
    \centering
        \includegraphics[width=0.6\textwidth]{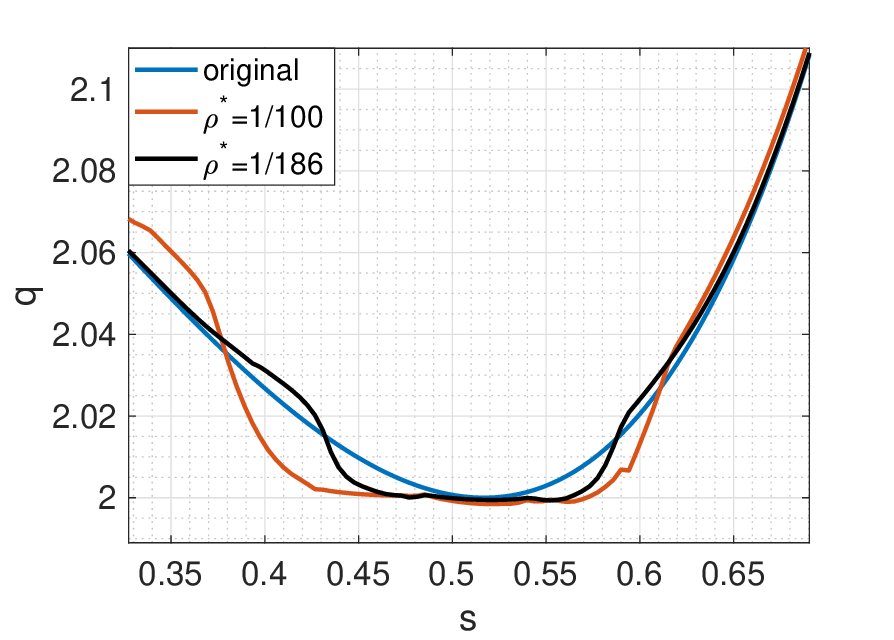}
       
    \caption{{Safety factor flattening for the $\rho^* =1/186$ and $ \rho^* = 1/100$ cases (black and red, respectively). Flux-driven EM simulations.}} 
    \label{rhostar_qmod}
\end{figure}

Concerning the flattening of the safety factor around the rational surface $q=2$, the $\rho^*=1/100$ simulation exhibits similar features to the CBC system size simulation. The comparison between the two is reported in Figure \ref{rhostar_qmod}. While they both exhibit the same qualitative behavior, it is interesting to see that the width of the flattened region  does depend on $\rho^*$, and it is indeed smaller for the CBC simulation (i.e. smaller region for smaller $\rho^*)$. Moreover, the differences are not left and right (with respect to $s=0.52$, i.e. the $q_{\text{min}}$ position) symmetric, but the effect of $\rho^*$ seems to be stronger in the left transport barrier.
 The region $s_{q_{min}} - s_{q_{2.01}}$ {is   $ \sim 14 \rho_i$ wide} for the CBC and $ \sim 10 \rho_i$ for the TCV-like case. On the right part ($s>s_{q_{\text{min}}}$) the region $s_{q_{2.01}} - s_{q_{\text{min}}} $ contains   $ \sim 11 \rho_i$ for the CBC and $ \sim 7 \rho_i$ for the TCV-like case. This is a further indication that even though the flattening is due to current spikes resulting from electron dynamics, the size of the flattened region is mediated by ion turbulence scale. 

It is interesting to note that flux-tube simulations that were modified to include q-profile radial variations of a given wavelength \cite{Ball_ppcf2023} did not find a dependence of the size of the flattened region on the q-profile wavelength \cite{Volcokas_Arxive2024_magn}.

\subsection{Sensitivity to $q_{\text {min}}$}

\begin{figure}[htbp]
    \centering
        \includegraphics[width=0.6\textwidth]{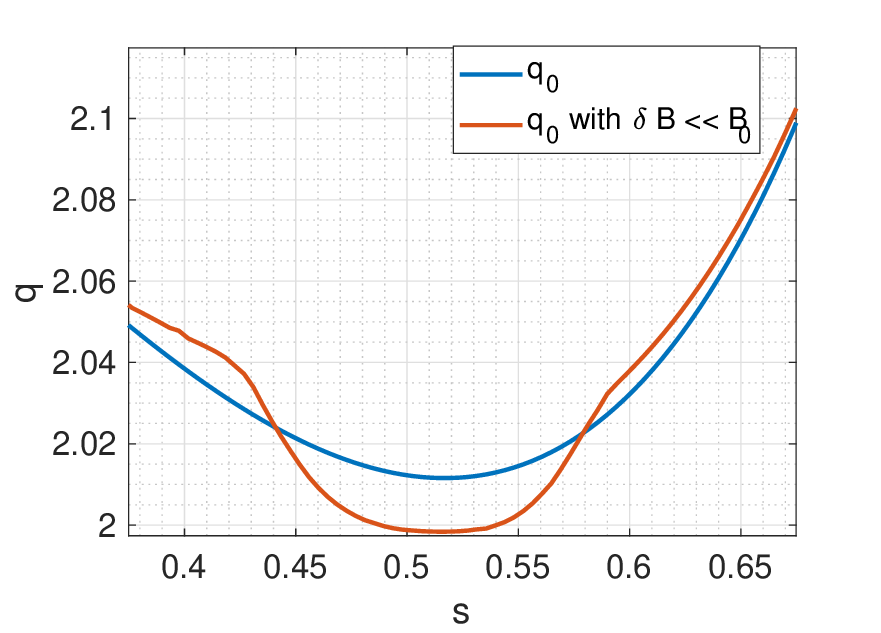}
       
    \caption{Initial and final q-profiles for the simulation with $\rho^*=1/100$ and initial $q_{\text {min}}=2.01$. Flux-driven EM simulation.} 
    \label{q_mod_201}
\end{figure}

\begin{figure}[htbp]
    \centering
    \begin{adjustbox}{addcode={}{},left}
        \hspace{-0cm}
        \includegraphics[width=1\textwidth]{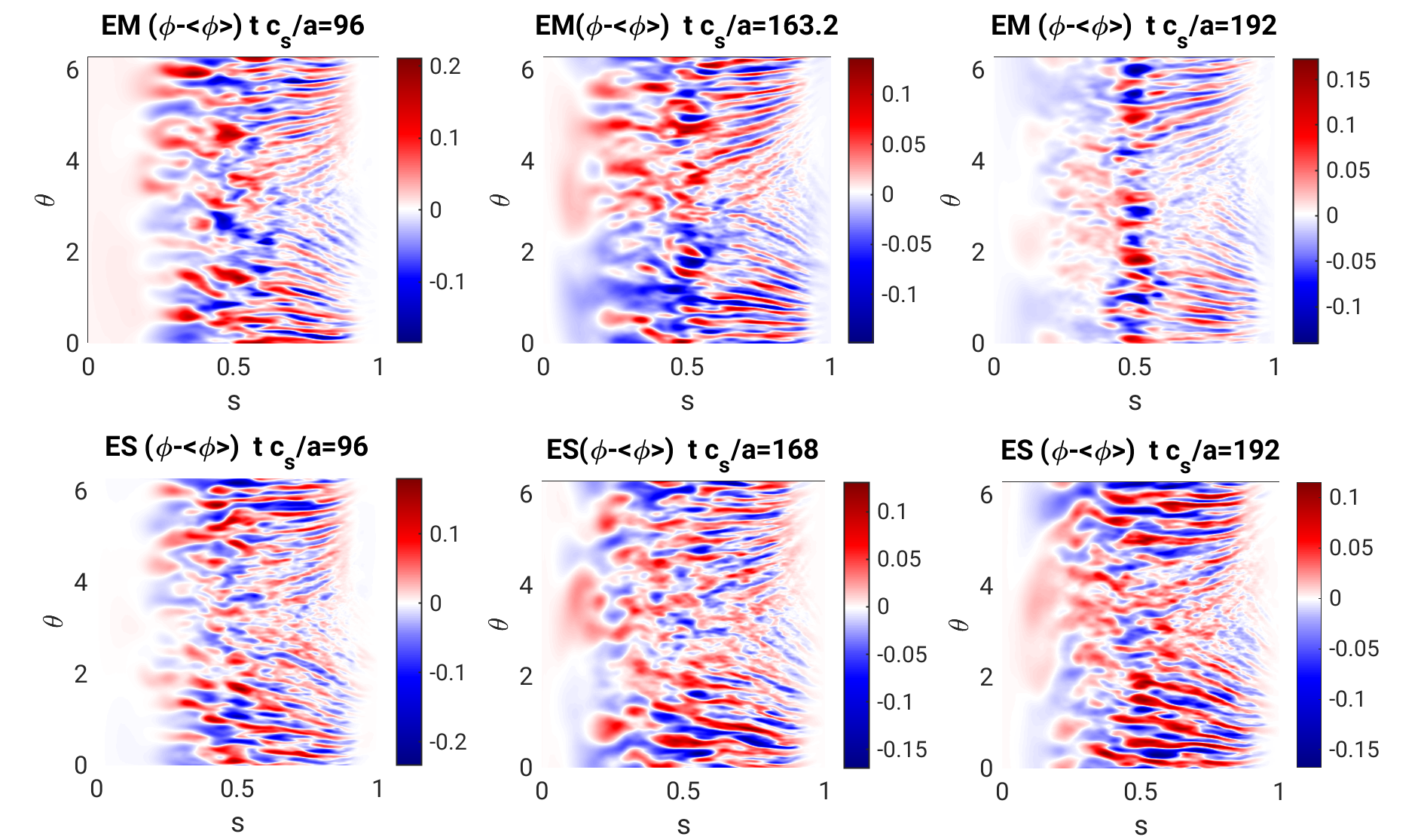}
        \begin{tikzpicture}[overlay,remember picture]
        \end{tikzpicture}
        
    \end{adjustbox}
    \caption{{Non-zonal electrostatic potential $\tilde{\phi} = \phi - \bar{\phi}$ (with $\bar{\phi}$ the zonal component) at the toroidal angle $\zeta=0$ for different times ($a/c_s $ normalization). Top: flux-driven EM  case; bottom:  flux-driven ES case.}}
    \label{phi_pert_ES_EM}
\end{figure}

\begin{figure}[htbp]
    \centering
        \includegraphics[width=1\textwidth]{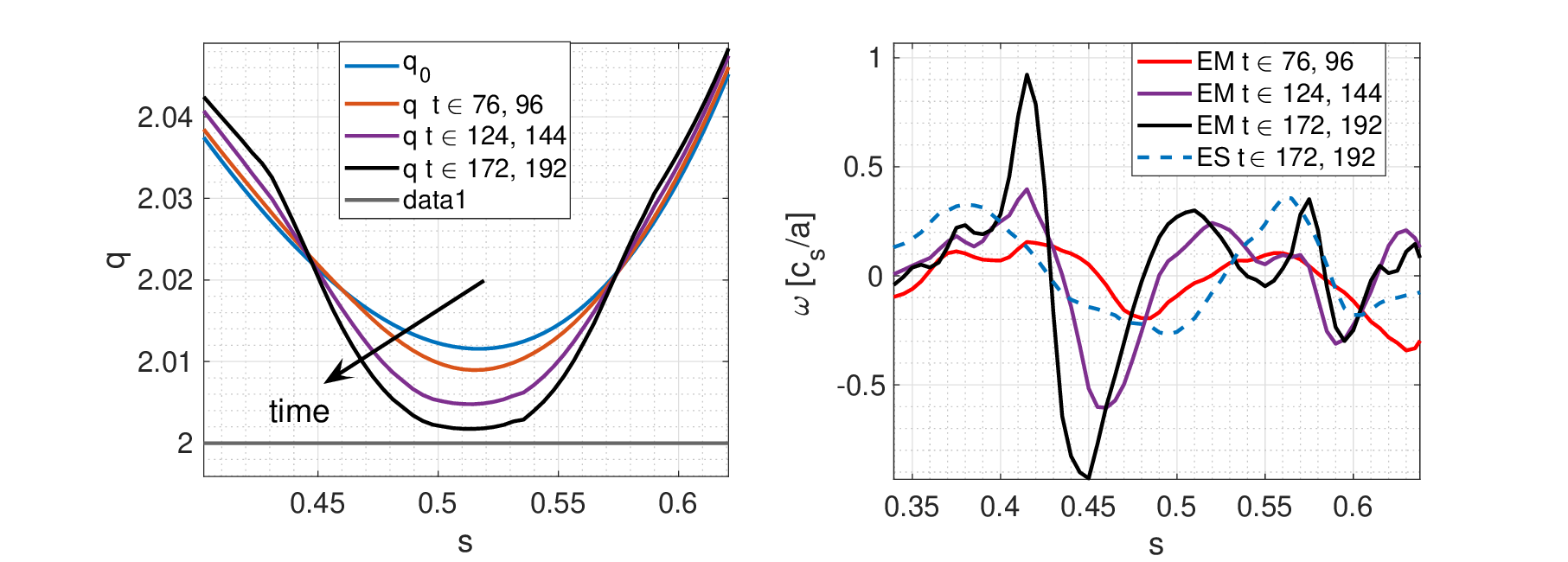}
       
    \caption{Time averaged radial profiles of safety factor (right) and ${\bf E} \times {\bf B} $ shearing rate $\langle \omega_{E \times B} \rangle_t$. The profiles for three different time windows are shown. { Flux-driven EM case with  initial $q_\text{min} = 2.01$ (in the right panel one time average of the flux-driven ES case is also shown as comparison). }}
    \label{time_trace_qmod}
\end{figure}

Another effect due to the $\rho^*$ parameter is the sensitivity of the system to $q_{\text {min}}$. Indeed, while for a rational $q_{\text {min}}$ there is a perfect self-interaction of turbulence, the self-interaction will happen also with  $q_{\text {min}}$ not perfectly rational, provided that the binormal size of the eddies is enough to allow for self-interaction. According to what is presented in \cite{Volcokas_Arxive2024}, the maximum excursion of $q_{\text {min}}$ that allows for self interaction is $\Delta q \sim \rho^* q_0/n$, with $q_o = m/n$ (i.e. $q=2/1$ in our case). According to this estimation, we expect for $\rho^* = 1/100$ that  $q_{\text {min}}  = 2.01 $ (i.e. $\Delta q = 0.01$) does still allow for some self interaction {(thus leading to similar effects to those described so far)}. This is confirmed by the simulations shown in Figure \ref{q_mod_201}. 
At this point it is interesting to assess the importance of EM effects. Indeed,  for this initial q-profile with $q_{\text {min}} =2.01$ the ES case will always have only a partial self interaction {(i.e. self-interaction without eddies exactly closing on themselves)}, while, thanks to the modification of q-profile, the EM case can have a self-interaction that is stronger after the turbulence modifies the safety factor: the smaller $\Delta q$ becomes  the more self-interaction one can expect. When self-interaction increases, we do expect to find similar features to those observed for the case $q_{\text {min}} =2$ (Figure \ref{phi_pert}-b). This is shown in Figure \ref{phi_pert_ES_EM}, where the quantity $\tilde{\phi} = \phi(s,\theta) - \bar{\phi}$ is shown for the ES case and for the EM case having both initially $q_{\text {min}} =2.01$. As one can see at the beginning they do exhibit  similar  $\tilde{\phi}$ features, but when the system evolves the EM and ES cases deviate. In particular, as it was for the case $q_{\text {min}} =2$, the fluctuations of the EM case become  uncorrelated around $s\simeq 0.45 $ and $s\simeq 0.58 $ while no substantial change in time is observed for the ES case. 
The change on the fluctuations correlation comes with the evolution of the safety factor and with the evolution of the zonal flows shearing rate, shown in Figure \ref{time_trace_qmod} averaged over $ [t_i - \Delta t, t_i] $ with $t_i \,c_s/a = 96, \, 144, \, 192$  and $\Delta t \, c_s/a=20$.


\begin{figure}[htbp]
    \centering
        \includegraphics[width=1\textwidth]{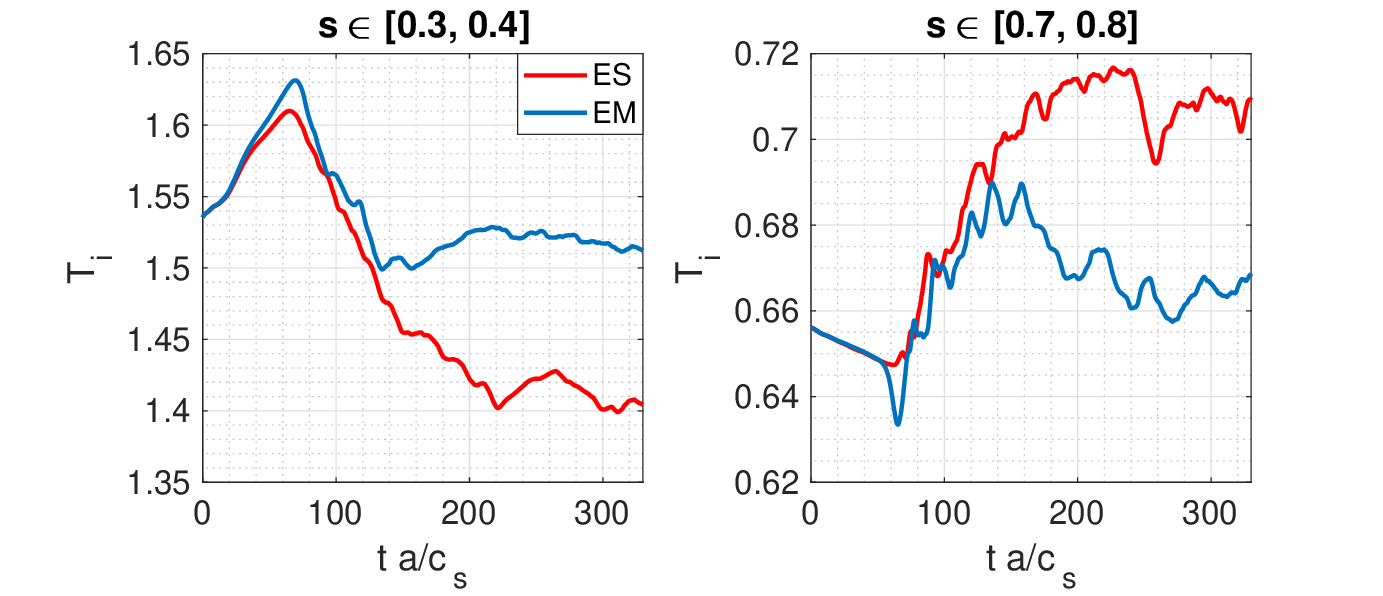}
       
    \caption{Temperature evolution in two radial windows, $s \in [0.3, 0.4]$ (left) and $s \in [0.7, 0.8]$ (right) for electrostatic and electromagnetic simulations, red and blue respectively. {Case with  initial $q_\text{min} = 2.01$. Flux-driven simulations.}}
    \label{temperature_evolution}
\end{figure}

To see the effect that the increased self-interaction has on the transport, the ion temperature evolution is displayed in Figure \ref{temperature_evolution}. Interestingly, the ES and the EM cases start to deviate from each other when the q profile starts to be modified and the strong $\omega_{E\times B}$ develops at $s \simeq 0.42$, coherently with the time windows displayed in Figure \ref{time_trace_qmod} (i.e. around $t \in [124, 144]$). 

\subsection{GD - FD comparison}
\begin{figure}[htbp]
    \centering
        \includegraphics[width=0.9\textwidth]{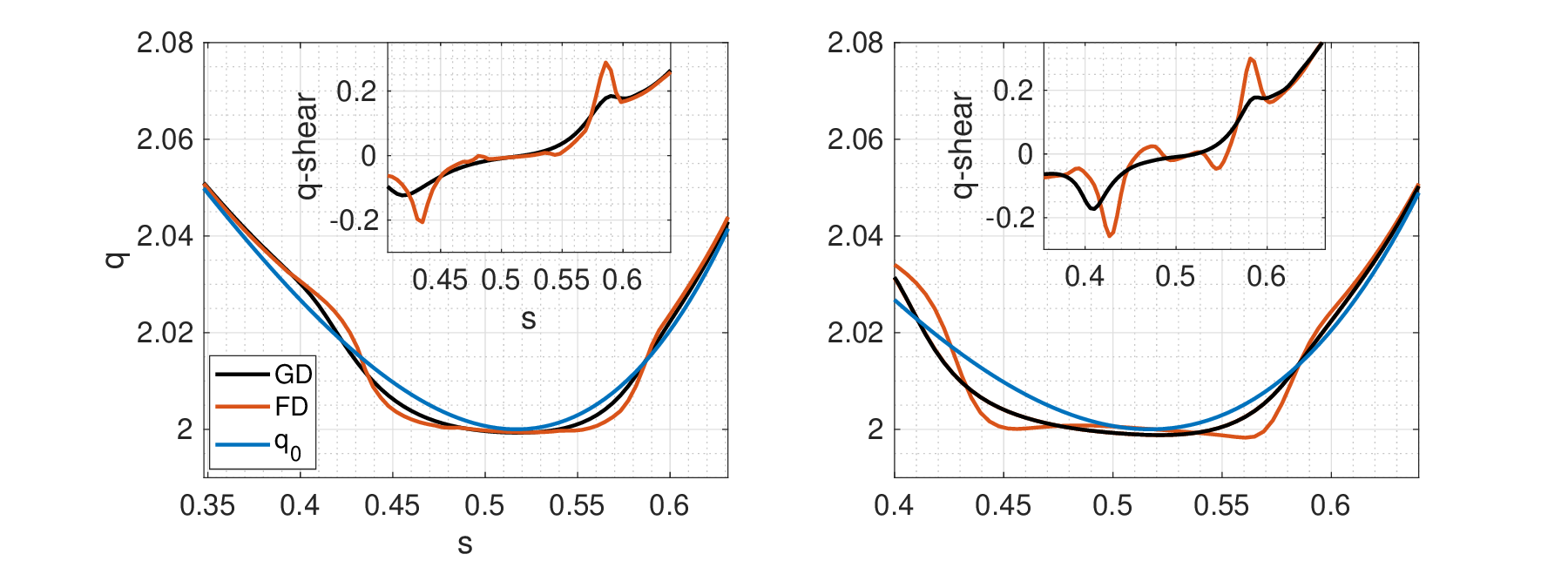}
        \caption{Safety factor flattening for the $\rho^* =1/186$ case for the gradient-driven (black) and flux-driven (red) simulations.  In the subplot the modified magnetic shear is shown. {Profiles computed over the time interval $t\,c_s/a \in [100, 200] $ (left) and over the time interval over the time interval $t\,c_s/a \in [200, 400] $ (right). EM simulations.}}
    \label{FD_GD_comparison_q}
\end{figure}

Finally, while any other comparison between flux-driven (FD) and gradient-driven (GD) simulations is difficult due to different methodologies, we can still compare how freely the system can evolve when subjected to some temperature profile constraint. One parameter that seems fair to compare is the q-profile flattening that results from turbulence. The comparison, {shown in Figure \ref{FD_GD_comparison_q}, is done  for two different time windows:  $t \, c_s/a \in [100,200]$ and $t \, c_s/a \in [200,400]$}. 
It is quite evident that the system evolves more freely in a flux-driven  run and the resulting q-profile flattens even more than the gradient-driven case, {although the difference is not dramatic and shows that the gradient-driven case is also able to capture most of the flattening. However, for the modified magnetic shear there is a strong difference at the edges of the flattened region}.  
 {While the direct effects of the local magnetic shear increase are not yet fully clear in our simulations, the flux-tube study conducted in Ref. \cite{Volcokas_Arxive2024_magn} suggests that its impact is secondary compared to non-linear effects brought by self-interaction}.

Indeed, this phenomenon comes along with high ZF shearing rates that are already very effective at  reducing  turbulent transport, both by non-linearly tearing apart eddies, and by linearly lowering growth rates of the ITG when $\omega_{E\times B} $ exceeds the growth rates \cite{JSama_Arxiv2024}.

\section{Conclusions and Outlook} \label{section_conclusions}

In this work we discussed the conditions that enable a strong reduction of heat transport and the development of an internal transport barrier (ITB) in a reversed-shear tokamak configuration, with the zero shear position that corresponds to a low-order rational surface, $q_{\text{min}}=2$. 
Our simulations setup is inspired by Cyclone Base Case parameters but excludes  density gradients.

In line with previous studies, we found that the adiabatic electron response assumption does not reveal the effect of the low-order 
 rational surface $q_{\text{min}}=2$. This is summarized in Figure \ref{non_linear}, where a linear trend is found in the transport coefficient with $q_{\text{min}}$ and no special sensitivity to $q_{\text{min}}=2$ is detected.

When a kinetic electron response is included, the system becomes very sensitive to $q_{\text{min}}=2$. {This is primarily due to kinetic electrons setting parallel eddy length and leading to strong parallel turbulence self-interaction \cite{Barnes_2011_PRL,Volcokas_nf2023}.} 
The first study has been performed employing the hybrid electron model (described in Section \ref{section_setup}) and the results are shown  Figure \ref{q_scan_hybrid}, where three different $q_{\text{min}}$ values are compared. We showed that when $q_{\text{min}}=2$ (namely it lies on a low order rational surface)  the heat transport drops by a factor of two in gradient-driven simulations compared to simulations where $q_{\text{min}} =2.5$ or even very close to 2, namely $q_{\text{min}}=2.03$. This  highlights that the heat flux dependence on the safety factor is very sharp. Remarkably, the strong reduction of heat transport is not limited to the radial location surrounding the rational surface  but it is spread all over the radial domain.

Using the fully-kinetic electron model, a good qualitative agreement is found with the hybrid model. Similar corrugations in the logarithmic gradients and zonal flow shearing rate $ \omega_{E\times B}$ are found (Figures \ref{hybrid_fullyk}-\ref{ExB_shear_fullykin_vs_hyb}) even though stronger peaking of the logarithmic gradients is observed using the fully-kinetic model.

We then analyzed the effects of a small $\beta$ ($\beta = 4 \;10^{-4}$). Consistently with flux-tube gyrokinetic findings, we found that the strong current layers due to electron dynamics develop in such a way as  to lead to an $A_\parallel$ that flattens out the q-profile around the zero shear position. This feedback happens only when  $q_{\text{min}}$ is sufficiently close to a rational value to enable  self-interaction of eddies. Three cases have been studied in this respect:  $q_{\text{min}} = 2,\; 2.01 \;\text{and}\; 2.03$. First we compared the cases $q_{\text{min}} = 2$ and $q_{\text{min}} = 2.03$: {the former has eddies biting their own tail and strong self-interaction, while the latter shows no strong self-interaction.}

This is shown in Figure \ref{parallel_elongation}, where the parallel elongation of eddies is high enough for them to bite their own tail in the  $q_{\text{min}} = 2$ case. This difference reflects on the final form of the safety factor profile, that gets modified for the $q_{\text{min}} = 2$ case while it does not change for $q_{\text{min}} = 2.03$. The new safety factor is shown in Figure \ref{rhostar_qmod} (for two different $\rho^*$). It is clear that there is a system size effect on the flattened area, with larger $\rho^*$ leading to a larger flattened area.
The case with $q_{\text{min}} = 2.01$ is different. Here $q_{\text{min}} $ is close enough to $2$ to enable self-interaction. This self-interaction results in a positive  feedback loop: it pushes the safety factor toward the rational value and this increases the self-interaction {which leads to further changes in the safety factor profile and turbulence dynamics}. This is shown in Figures \ref{q_mod_201},\ref{time_trace_qmod} {with the resulting effect on fluctuations shown in Figure \ref{phi_pert_ES_EM}}.

The most important result of this work is the demonstration of the formation of an ITB; we compared two similar simulations, one clearly shows an ITB and the other one does not.  Specifically, through flux-driven simulations we have shown that  by tailoring $q_{\text{min}}$ it is possible to achieve an ITB with high ion temperature on axis.  We compared the evolution of the $q_{\text{min}} = 2$ case with the evolution of the $q_{\text{min}} = 2.03$ case. The final profiles of the two simulations are shown  in Figure \ref{comparison_qmin2_qbar2}, proving that the transport barrier is lost when $q_{\text{min}} = 2.03$.

Regarding  transport scaling, this regime seems to be close to GyroBohm: considering the same plasma conditions ($T, n, B$)  going from a TCV-like $\rho^*$ ($\rho^* = 1/100 $ at $s=0.52$) to DIII-D like $\rho^*$ ($\rho^* = 1/186 $ at $s=0.52$) requires only $20\%$ more power, and it even features larger gradients (thus larger temperatures on axis).

{The sensitivity to $q_{\text{min}}$ for transport barrier formation does depend on $\rho^*$. We tested   $q_{\text{min}} = 2.01$ for $\rho^*=1/100$ and in this case we still observe the emergence of transport barrier {as this  $q_{\text{min}} $ allows for turbulence self-interaction}. Particularly interesting is the fact that with this  $q_{\text{min}} $ the system starts to develop the transport barrier  when self-interaction  increases (i.e. q-profile is dragged towards  $q_{\text{min}} = 2$).}

Comparing electrostatic and electromagnetic simulations, we found that the flattening of the safety factor per se is not a strongly stabilizing mechanism when self-interaction is already maximized (i.e. $q_{\text{min}} = 2$), but it plays an important role when self-interaction is partial: indeed significant differences  appear between  the electrostatic and electromagnetic cases for $q_{\text{min}} = 2.01$ (see Figure \ref{temperature_evolution}).

\section{Acknowledgments}

The authors thank  {B. Rofman}  for fruitful discussions, E. Lanti for its continuous support to the ORB5 code and {Alexey Mishchenko for his support to this work providing additional  computational time}.

This work has been carried out within the framework of the EUROfusion Consortium, partially funded by the European Union via the Euratom Research and Training Programme (Grant Agreement No 101052200 — EUROfusion). The Swiss contribution to this work has been funded by the Swiss State Secretariat for Education, Research and Innovation (SERI). Views and opinions expressed are however those of the author(s) only and do not necessarily reflect those of the European Union, the European Commission or SERI. Neither the European Union nor the European Commission nor SERI can be held responsible for them.
This work is also supported by a grant from the Swiss National Supercomputing Centre
(CSCS) under projects ID s1249, s1252 and was partly supported by the Swiss National Science Foundation.

\appendix
\section{Assessment of importance of the term $\text{d}f_0/\text{d}t \big|_0$} \label{A_1}
\begin{figure}[htbp]
    \centering
    \begin{adjustbox}{addcode={}{},left}
        \hspace{-1.5cm}
        \includegraphics[width=1.2\textwidth]{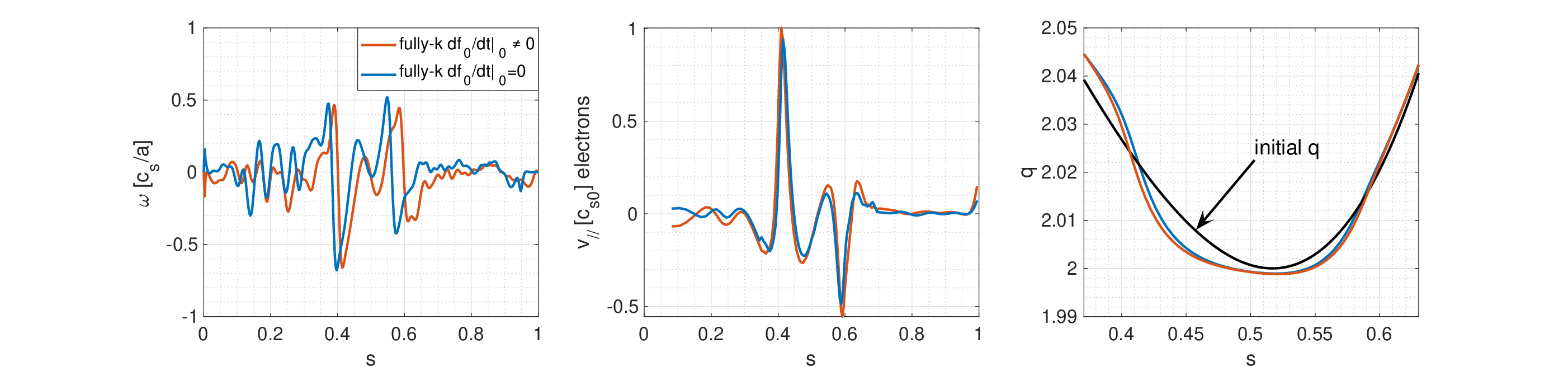}
        \begin{tikzpicture}[overlay,remember picture]
        \end{tikzpicture}
        
    \end{adjustbox}
    \caption{{Comparison between simulations with (red) and without (blue) the $\text{d}f_0/\text{d}t \big|_0 $ term. Time averaged radial profiles of  ${\bf E} \times {\bf B} $ shearing rate $\langle \omega_{E \times B} \rangle_t$ (left), electron $v_\parallel$ (central), $q$-flattening (right). The blue case is the same case presented in Section \ref{EM_GD}. EM simulations.}}
    \label{E_time_B_and_q_neoclas}
\end{figure}

{In this appendix we assess the importance of   including the term $\text{d}f_0/\text{d}t \big|_0$ into the dynamics. Whereas  we only focus on the effects on our numerical setup, the importance of coherently including $\text{d}f_0/\text{d}t \big|_0$   is widely discussed in ref \cite{Angelino_2006_pop}. In most GK simulations the distribution function is split as  $f = f_0 + \delta f$, with $f_0$  the initial distribution function (or evolving control variate in the case of background adaptation scheme). Neglecting in this discussion the fact the one could adapt $f_0$ during the simulation (with the adaptation being anyway limited in a gradient-driven simulation), the common approach is to shape $f_0$ as a Maxwellian function of temperature and density at a specific radial coordinate and to consider it an equilibrium for the unperturbed collisionless system (thus $\text{d}f_0/\text{d}t \big|_0 = 0$ by definition). However,  {taking $f_0$ constant on a magnetic surface does not make it an equilibrium for collisionless gyrokinetic (not even for collisional one)}, making  $\text{d}f_0/\text{d}t \big|_0 = 0$ an approximation. Indeed in toroidal axisymmetric system, the unperturbed orbits are characterized by three constants of motion; the particle energy,  magnetic moment and  toroidal canonical momentum $\psi_0$:
\begin{equation}
\psi_0 = \psi + \frac{q_s \, B_\phi(\psi)}{m_s \, B(\psi)} \, v_\parallel, 
\end{equation}

Thus, since the Maxwellian distribution function used in most works is defined as function of $\psi$ (local Maxwellian) and not $\psi_0$ (canonical Maxwellian), the initial distribution function $f_0$ is not an equilibrium along the unperturbed orbits. We stress that while in flux-tube is not possible to use $\psi_0$ to construct $f_0$, in global simulations this approach can be followed \cite{Angelino_2006_pop}.

In this work for simplicity we used a local Maxwellian and, for the fully kinetic electrons simulations, we used the approximation  $\text{d}f_0/\text{d}t \big|_0 = 0$ to speed-up the computations. Thus, for the fully-kinetic electrons simulations we are neglecting a part of the dynamics that is present. We thus performed an additional temperature-gradient-driven, electromagnetic ($\beta=4 \,10^{-4}$) fully-kinetic electron model simulation (the same numerical setup of the simulation presented in Section \ref{EM_GD} with $q_{\text{min}}=2$) including the $\text{d}f_0/\text{d}t \big|_0 $ term and compared the output with the twin simulation that does not include the $\text{d}f_0/\text{d}t \big|_0 $ term.

The result is shown in Figure \ref{full_f} where we use $\chi$   as proxy for the comparison. 
From the plot one can see that including the $\text{d}f_0/\text{d}t \big|_0 $ term does not change the main points and conclusions of our work. If anything, the barrier seems even stronger with a smaller $\chi$ around $s=0.6$. The larger difference between the the simulations is in the very core. Since we have seen the turbulence is stabilized at $s < 0.4$, this difference must come from the additional term. To stress the importance of the $\text{d}f_0/\text{d}t \big|_0 $ term, we plot in Figure \ref{neo_only} the contribution to $\chi$ coming from the magnetic drifts only ($\chi_{B}$ for simplicity of notation). For a further check, we have also inserted for comparison the hybrid case. One can see indeed that when evaluating the magnetic drifts contributions only, the $\chi_{B}$ of the fully-kinetic case including the $\text{d}f_0/\text{d}t \big|_0 $ term is much more similar to the hybrid $\chi_{B}$ (that does also includes the $\text{d}f_0/\text{d}t \big|_0 $ term) than to the other fully-kinetic case that does not include the $\text{d}f_0/\text{d}t \big|_0 $ term.

\begin{figure}[htbp]
    \centering
        \includegraphics[width=0.8\textwidth]{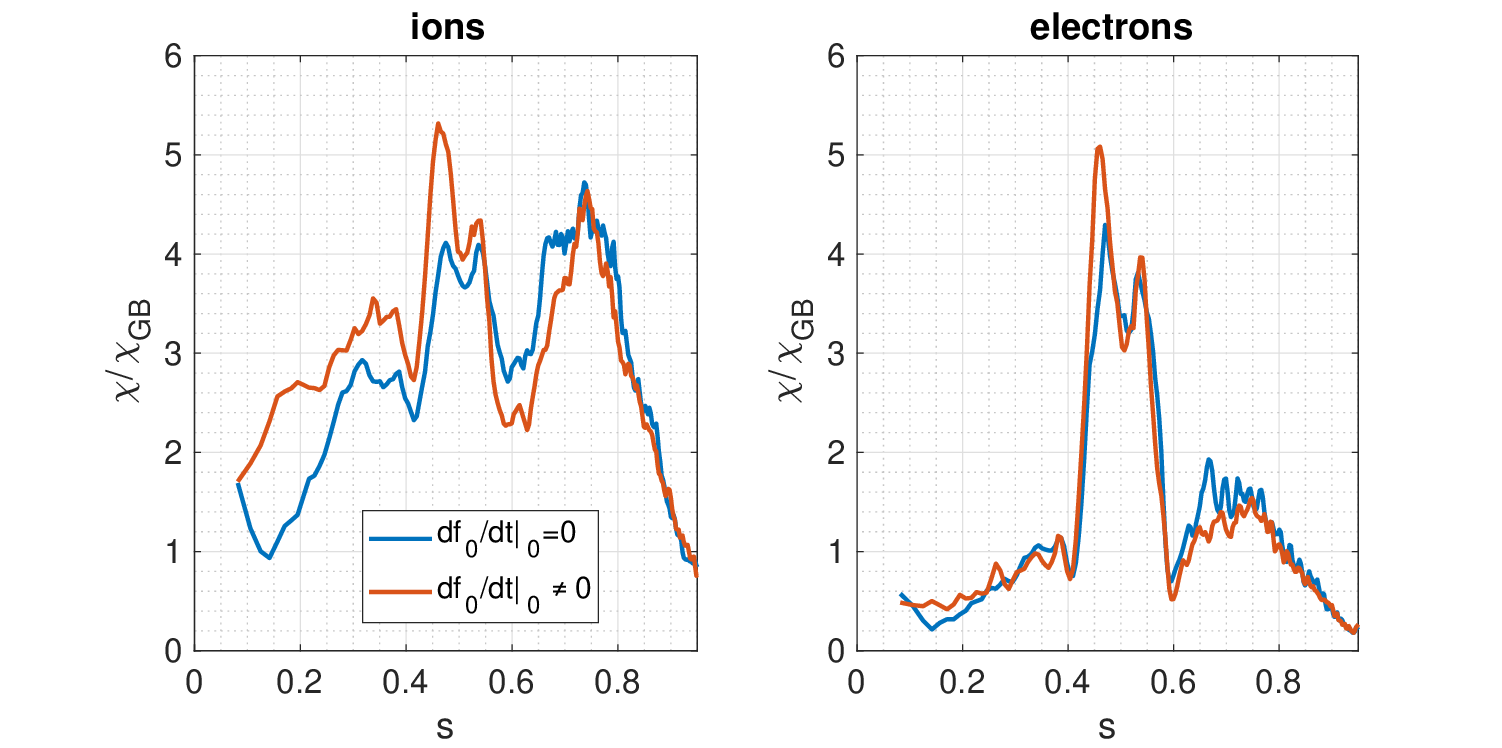}
        \begin{tikzpicture}[overlay,remember picture]
        \end{tikzpicture}
        
    \caption{{Time averaged radial profiles of  $\chi$ for simulations with (red) and without (blue) the $\text{d}f_0/\text{d}t \big|_0 $ term. The blue case is the same case presented in Section \ref{EM_GD}.  Gradient-driven EM simulations. }}
    \label{full_f}
\end{figure}

\begin{figure}[htbp]
    \centering
        \includegraphics[width=0.8\textwidth]{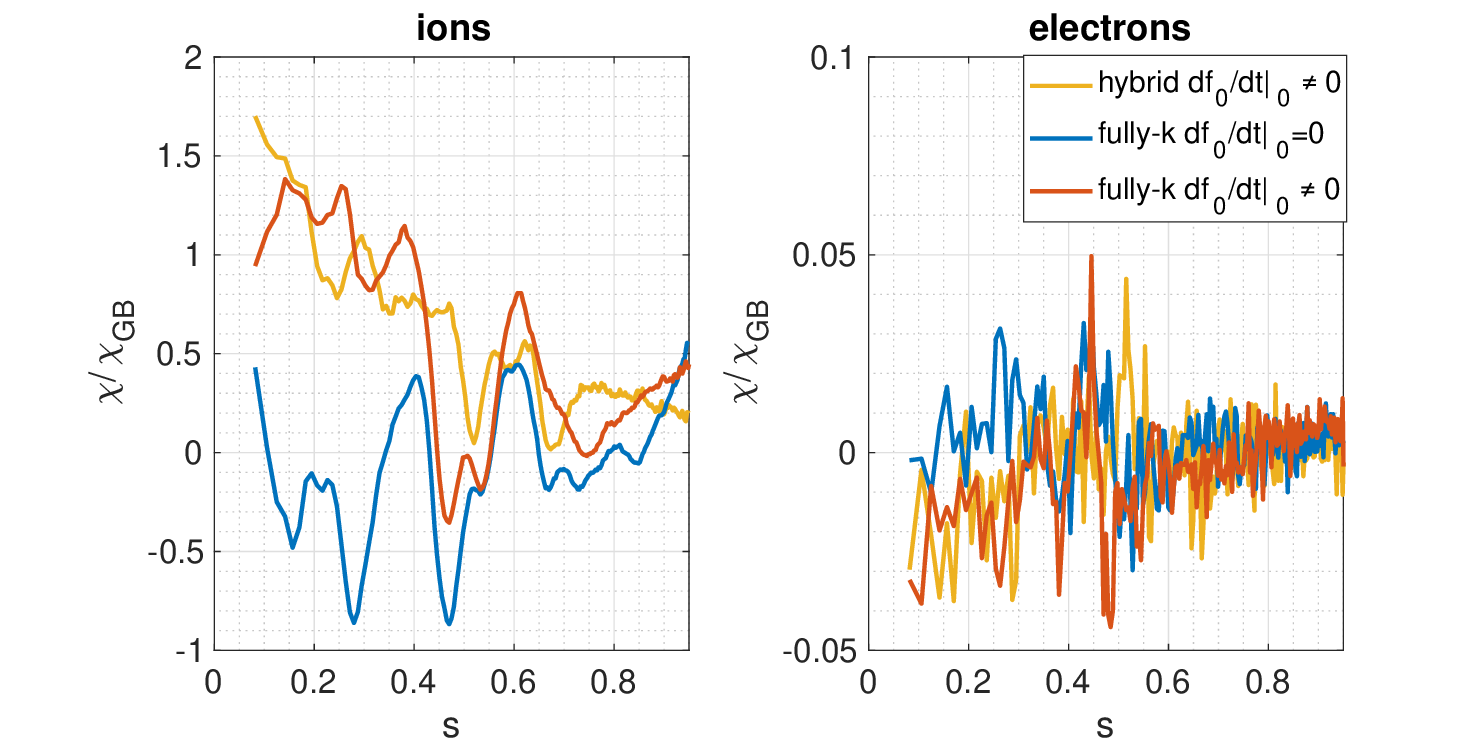}
        \begin{tikzpicture}[overlay,remember picture]
        \end{tikzpicture}
        
    \caption{{ Time averaged radial profiles of  $\chi$ ($\chi$ computed including the fluxes due to magnetic drifts only) for simulations with (red-yellow) and without (blue) the $\text{d}f_0/\text{d}t \big|_0 $ term. EM fully-kinetic electron  (red, blue) and ES hybrid  (yellow) models are used. The blue case is the same case presented in Section \ref{EM_GD}, the yellow case is the same case presented in Section \ref{hybrid_section_results}. Gradient-driven simulations.}}
    \label{neo_only}
\end{figure}

The other crucial quantities to compare for the relevance of our studies are the time-averaged ${\bf E} \times {\bf B} $ shearing rate $\langle \omega_{E \times B} \rangle_t$ and the structure of the electron parallel velocity that we have seen to lead to $q$-flattening. These quantities are shown in Figure \ref{E_time_B_and_q_neoclas}. With this comparison we can really discriminate between the different spatial scales that affect the stationary  ${\bf E} \times {\bf B} $ shearing rate and the generation of the currents leading to $q$-flattening. For a given equilibrium, the difference between $\psi$ and $\psi_0$ goes with $v_\parallel/m$ thus it is much larger for the ions than for the electrons. The fact that the  ${\bf E} \times {\bf B} $ shearing rate (Figure \ref{E_time_B_and_q_neoclas}-left) has a detectable radial shift of about $\Delta r_{peak}\simeq 9 \rho_i$ when the $\text{d}f_0/\text{d}t \big|_0 $ term is included means that this quantity depends on the ion dynamics (yet we have seen that also including non-adiabaticity of electrons is crucial). On the other hand, the electron $v_\parallel$ is very similar and so is the resulting $q$-flattening (Figure \ref{E_time_B_and_q_neoclas}-central,right panels), meaning that this mechanism is essentially tied to the electron dynamics and seems to be unaffected by the additional term on the ion dynamics.}

\vspace*{2cm}
\bibliographystyle{unsrt}
\bibliography{Bibliography}

\end{document}